\documentclass[11pt]{article}
\usepackage{graphicx,amsmath,amssymb}

\begin{document}

\newtheorem{observation}{Observation}
\newtheorem{theorem}{Theorem}[section]
\newtheorem{lemma}[theorem]{Lemma}
\newtheorem{corollary}[theorem]{Corollary}
\newtheorem{proposition}[theorem]{Proposition}
\newcommand{\blackslug}{\penalty 1000\hbox{
    \vrule height 8pt width .4pt\hskip -.4pt
    \vbox{\hrule width 8pt height .4pt\vskip -.4pt
          \vskip 8pt
      \vskip -.4pt\hrule width 8pt height .4pt}
    \hskip -3.9pt
    \vrule height 8pt width .4pt}}
\newcommand{\proofend}{\quad\blackslug}
\newenvironment{proof}{$\;$\newline \noindent {\sc Proof.}$\;\;\;$\rm}{\qed}
\newcommand{\qed}{\hspace*{\fill}\blackslug}
\newenvironment{definition}{$\;$\newline \noindent {\bf Definition}$\;$}{$\;$\newline}
\def\boxit#1{\vbox{\hrule\hbox{\vrule\kern4pt
  \vbox{\kern1pt#1\kern1pt}
\kern2pt\vrule}\hrule}}
\addtolength{\baselineskip}{+0.4mm}

\bibliographystyle{elsarticle-num}

\title{\vspace{-.4in} \bf
Parameterized Algorithms for\\
           the Maximum Agreement Forest Problem\\
           on Multiple Rooted Multifurcating Trees}


\author{
 \vspace*{3mm}
 {\sc Feng Shi}$^{\mbox{\footnotesize \S}}$ \ \ \ \
 {\sc Jianer Chen}$^{\mbox{\footnotesize \S\textdaggerdbl}}$ \ \
  {\sc Qilong Feng}$^{\mbox{\footnotesize \S}}$ \ \
 {\sc Jianxin Wang}$^{\mbox{\footnotesize \S}}$\footnote{Corresponding author, email:
 jxwang@mail.csu.edu.cn} \\ \\
 $^{\mbox{\footnotesize \S}}$School of Information Science and Engineering \\
   Central South University \\
  \vspace*{3mm}
   Changsha 410083, P.R. China\\
   $^{\mbox{\footnotesize \textdaggerdbl}}$Department of Computer Science and Engineering\\
   Texas A\&M University\\
   College Station, Texas 77843, USA  }

\date{}
\maketitle

\vspace{-7mm}

\begin{abstract}
The Maximum Agreement Forest problem has been extensively studied in phylogenetics. Most
previous work is on two binary phylogenetic trees. In this paper, we study a generalized
version of the problem: the Maximum Agreement Forest problem on multiple rooted multifurcating
phylogenetic trees, from the perspective of fixed-parameter algorithms. By taking advantage
of a new branch-and-bound strategy, two parameterized algorithms, with running times
$O(2.42^k m^3 n^4)$ and $O(2.74^k m^3 n^5)$, respectively, are presented for the hard
version and the soft version of the problem, which correspond to two different biological
meanings to the polytomies in multifurcating phylogenetic trees.
\end{abstract}

\section{Introduction}

Phylogenetic trees (alternatively called evolutionary trees) are an invaluable tool in phylogenetics that are used to represent the evolutionary histories of homologous regions of genomes from a collection of extant species or, more generally, taxa. However, due to reticulation events, such as hybridization, recombination, or lateral gene transfer (LGT) in evolution, phylogenetic trees constructed by different regions of genomes may have different structures. Since the reticulation events can be studied by examining these differences in structures, several metrics, such as Robinson-Foulds distance \cite{1}, Nearest Neighbor Interchange (NNI) distance \cite{2}, Hybridization number \cite{hybridization}, Tree Bisection and Reconnection (TBR) distance, and Subtree Prune and Regraft (SPR) distance \cite{3,4}, have been proposed in the literature to compare these different phylogenetic trees. Among these metrics, the SPR distance has been studied extensively for investigating phylogenetic inference \cite{dudas}, lateral genetic transfer \cite{beiko,whidden5}, and MCMC search \cite{whidden6}.

Given two phylogenetic trees on the same collection of taxa, the SPR distance between the two trees is defined to be the minimum number of ``Subtree Prune and Regraft'' operations \cite{6} needed to convert one tree to the other. Since the Subtree Prune and Regraft operation has been widely used as a method to model a reticulation event, the SPR distance provides a lower bound on the number of reticulation events needed to reconcile the two phylogenetic trees \cite{10}, which can give an indication how reticulation events influence the evolutionary history of the taxa under consideration.

For the study of SPR distance, Hein {\it et al.}~\cite{5} proposed the concept of {\it maximum agreement forest} (MAF) for two phylogenetic trees, which is a common subforest of the two trees with the minimum {\it order} among all common subforests of the two trees (the order of a forest is defined as the number of connected components of the forest). Bordewich and Semple \cite{7} proved that the order of an MAF for two {\it rooted binary} phylogenetic trees minus $1$ is equal to their rSPR distance. Since then, much work has been focused on studying the Maximum Agreement Forest problem on two rooted binary phylogenetic trees, which asks for an MAF for the two trees.

Biological researchers traditionally assumed that phylogenetic trees were bifurcating \cite{21,22}, which motivated most earlier work focused on the Maximum Agreement Forest problem for binary trees. However, more recent research in biology and phylogenetics has called a need to study the problem for general trees. For example, for many biological data sets in practice \cite{27,28}, the constructed phylogenetic trees always contain {\it polytomies} (alternatively called {\it multifurcations}). There are two different meanings to the polytomies in phylogenetic trees: (1) the polytomy refers to an event during which an ancestral species gave rise to more than two offspring species at the same time \cite{23,24,25,26}, which is called a {\it hard} polytomy; (2) the polytomy refers to ambiguous evolutionary relationships as a result of insufficient information, which is called a {\it soft} polytomy. Note that the types of polytomies in the phylogenetic trees have a substantial impact on designing algorithms for comparing these trees. For example, a soft polytomy with three leaves $(a,b,c)$ is not considered different from two resolved bifurcations of the same three leaves $((a,b),c)$, as the soft polytomy is ambiguous rather than conflicting, and the soft polytomy $(a,b,c)$ can be {\it binary resolved} as $((a,b),c)$. On the other hand, if the polytomy $(a,b,c)$ is hard, then $(a,b,c)$ and $((a,b),c)$ are considered different as the hard polytomy is interpreted as simultaneous speciation. In this paper, we study two versions of the Maximum Agreement Forest problem on rooted multifurcating trees: (1) the hard version, which assumes that all polytomies in the multifurcating phylogenetic trees are hard; and (2) the soft version, which assumes that all polytomies in the multifurcating phylogenetic trees are soft.

Because of the two types of polytomies, two types of rSPR distance are defined. Given two rooted multifurcating phylogenetic trees $T_1$ and $T_2$, the {\it hard rSPR distance} between $T_1$ and $T_2$ is defined as the minimum number of rSPR operations needed to transform one tree into the other under the assumption that all polytomies in the two trees are hard \footnote{The relationship between MAF and the metric of rSPR distance on binary trees can be naturally extended to that on multifurcating trees~\cite{17},~\cite{14}.}, and the {\it soft rSPR distance} between $T_1$ and $T_2$ is defined as the minimum rSPR distance between all pairs of {\it binary resolutions} of $T_1$ and $T_2$~\cite{17}. Apparently, the hard rSPR distance captures all structural differences between the two trees, and the soft rSPR distance only captures the structural differences that cannot be reconciled by resolving the multifurcations appropriately. The hard rSPR distance between two multifurcating phylogenetic trees corresponds to their MAF under the assumption that all polytomies are hard, and the soft rSPR distance between two multifurcating phylogenetic trees corresponds to their MAF under the assumption that all polytomies are soft.

For the same collection of taxa, multiple (i.e., two or more) different phylogenetic trees may be constructed based on different data sets or different building methods. Studying the Maximum Agreement Forest problem on multiple phylogenetic trees has more biological meaning than that on two trees. For example, suppose that we have two phylogenetic trees that are constructed by two homologous regions of genomes from a collection of taxa. As mentioned above, studying the order of their MAF can indicate how reticulation events influence the evolutionary histories of two homologous regions of the genomes. Note that these reticulation events that influenced the evolutionary histories of the two homologous regions of the genomes may also influence the evolutionary histories of other homologous regions of the genomes. Thus, if we construct phylogenetic trees for each homologous region of the genomes, and study their MAF, then the order of their MAF can give a more comprehensive indication of the extent to which reticulation has influenced the evolutionary history of the collection of taxa. Moreover, consider an MAF $F$ (hard version or soft version) of order $k$ for a set $\mathcal{C}$ of rooted phylogenetic trees. Since $F$ is also an agreement forest (not necessarily an MAF) for any two trees $T_i$ and $T_j$ in $\mathcal{C}$, the (hard or soft) rSPR distance between $T_i$ and $T_j$ would not be greater than $k-1$. Thus, the order of an MAF for $\mathcal{C}$ provides an upper bound for the rSPR distance between any two trees in $\mathcal{C}$. Last but not least, constructing an MAF for multiple phylogenetic trees is a critical step in studying the reticulate networks with the minimum number of reticulation vertices for multiple phylogenetic trees \cite{18}, which is a hot topic in phylogenetics. The reason is that among all reticulate networks for the given multiple phylogenetic trees, the number of reticulation vertices in the reticulate network with the minimum number of reticulation vertices is equal to the order of an MAF for the given multiple phylogenetic trees minus one if the MAF is acyclic~\cite{leo2013}.

To summarize, it makes perfect sense to study the Maximum Agreement Forest problem on multiple rooted multifurcating phylogenetic trees. In this paper, we will focus on parameterized algorithms for the two versions (the hard version and the soft version) of the Maximum Agreement Forest problem on multiple rooted multifurcating phylogenetic trees. In the following, we first review previous related work on the Maximum Agreement Forest problem. Note that there are two kinds of phylogenetic trees, rooted or unrooted. The only distinction between the two kinds of phylogenetic trees is that whether an ancestor-descendant relation is defined in the tree. Although in this paper we only study the rooted phylogenetic trees, we also present previous related work on  unrooted phylogenetic trees. In particular, Allen and Steel \cite{6} proved that the TBR distance between two unrooted binary phylogenetic trees is equal to the order of their MAF minus $1$.

In terms of the computational complexity of the problems, it has been proved that computing the order of an MAF is NP-hard and MAX SNP-hard for two unrooted binary phylogenetic trees \cite{5}, as well as for two rooted binary phylogenetic trees \cite{7}.

\smallskip

{\bf Approximation Algorithms}. For the Maximum Agreement Forest problem on two rooted binary phylogenetic trees, Hein {\it et al.}~\cite{5} proposed an approximation algorithm of ratio $3$. However, Rodrigues {\it et al.}~\cite{8} found a subtle error in \cite{5}, showed that the algorithm in \cite{5} has ratio at least $4$, and presented a new approximation algorithm which they claimed has ratio $3$. Borchwich and Semple~\cite{7} corrected the definition of an MAF for the rSPR distance. Using this definition, Bonet {\it et al.}~\cite{9} provided a counterexample and showed that, with a slight modification, both the algorithms in \cite{5} and \cite{8} compute a $5$-approximation of the rSPR distance between two rooted binary phylogenetic trees in linear time. The approximation ratio was improved to 3 by Bordewich {\it et al.}~\cite{10}, but the running time of the algorithm is increased to $O(n^5)$. A second 3-approximation algorithm presented in \cite{11} achieves a running time of $O(n^2)$. Whidden {\it et al.}~\cite{12} presented the third 3-approximation algorithm, which runs in linear-time. Shi {\it et al.}~\cite{shiapprox} improved the ratio to 2.5, but the algorithm has running time $O(n^2)$. Recently, Schalekamp~\cite{Schalekamp} presented a 2-approximation algorithm by LP Duality (the running time is polynomial, but the exact order of the running time is not clear), which is the best known approximation algorithm for the Maximum Agreement Forest problem on two rooted binary trees. For the Maximum Agreement Forest problem on two unrooted binary phylogenetic trees, Whidden {\it et al.}~\cite{12} presented a linear-time approximation algorithm of ratio $3$, which is currently the best algorithm for the problem.

There are also several approximation algorithms for the Maximum Agreement Forest problem on two multifurcating phylogenetic trees. For the Maximum Agreement Forest problem on two rooted multifurcating phylogenetic trees, Rodrigues {\it et al.}~\cite{11} developed an approximation algorithm of ratio $d+1$ for the hard version, with running time $O(n^2 d^2)$, where $d$ is the maximum number of children a node in the input trees has. Lersel {\it et al.}~\cite{33} presented a 4-approximation algorithm with polynomial running time for the soft version. Recently, Whidden {\it et al.}~\cite{17} gave an improved $3$-approximation algorithm with running time $O(n \log n)$ for the soft version. For the Maximum Agreement Forest problem on two unrooted multifurcating phylogenetic trees, Chen {\it et al.}~\cite{14} developed a $3$-approximation algorithm with running time $O(n^2)$ for the hard version.

For the Maximum Agreement Forest problem on multiple rooted binary phylogenetic trees, Chataigner \cite{15} presented a polynomial-time approximation algorithm of ratio 8. Recently, Mukhopadhyay and Bhabak~\cite{Asish} and Chen {\it et al.}~\cite{20}, independently, developed two 3-approximation algorithms. The running times of the two algorithms in~\cite{Asish} and~\cite{20} are $O(n^2 m^2)$ and $O(n m \log n)$ respectively, where $n$ denotes the number of leaves in each phylogenetic tree, and $m$ denotes the number of phylogenetic trees in the input instance. For the Maximum Agreement Forest problem on multiple unrooted binary trees, Chen {\it et al.}~\cite{20} presented a $4$-approximation algorithm with running time $O(n m \log n)$ . To our best knowledge, there is no known approximation algorithm for the Maximum Agreement Forest problem on multiple rooted (unrooted) multifurcating phylogenetic trees.

\smallskip

{\bf Parameterized Algorithms}. Parameterized algorithms for the Maximum Agreement Forest problem, parameterized by the order $k$ of an MAF, have also been studied. A parameterized problem is {\it fixed-parameter tractable} \cite{fptbook} if it is solvable in time $f(k)n^{O(1)}$, where $n$ is the input size and $f$ is a computable function only depending on the parameter $k$. For the Maximum Agreement Forest problem on two unrooted binary phylogenetic trees, Allen and Steel \cite{6} showed that the problem is fixed-parameter tractable. Hallett and McCartin \cite{10} developed a parameterized algorithm of running time $O(4^k k^5+n^{O(1)})$ for the Maximum Agreement Forest problem on two unrooted binary phylogenetic trees. Whidden and Zeh \cite{12} further improved the time complexity to $O(4^k k + n^3)$. For the Maximum Agreement Forest problem on two rooted binary phylogenetic trees, Bordewich {\it et al.}~\cite{10} developed a parameterized algorithm of running time $O(4^k k^4 + n^3)$. Whidden {\it et al.}~\cite{siam} improved this bound and developed an algorithm of running time $O(2.42^k k + n^3)$. Chen {\it et al.}~\cite{zhizhongchen} presented an algorithm of running time $O(2.344^k n)$, which is the best known result of the Maximum Agreement Forest problem on two rooted binary phylogenetic trees.

There are also several parameterized algorithms for the Maximum Agreement Forest problem on two multifurcating phylogenetic trees. Whidden {\it et al.}~\cite{17} presented an algorithm of running time $O(2.42^k k + n^3)$ for the soft version of the Maximum Agreement Forest problem on two rooted multifurcating phylogenetic trees. Shi {\it et al.}~\cite{Yang} presented an algorithm of running time $O(4^k n^5)$ for the hard version of the Maximum Agreement Forest problem on two unrooted multifurcating phylogenetic trees. Chen {\it et al.}~\cite{14} developed an improved algorithm of running time $O(3^k n)$, which is the best known result for the hard version of the Maximum Agreement Forest problem on two unrooted multifurcating phylogenetic trees.

For the Maximum Agreement Forest problem on multiple rooted binary phylogenetic trees, Chen {\it et al.}~\cite{18} presented a parameterized algorithm of running time $O^*(6^k)$ \footnote{The $O^*$ notation means the polynomial factors of the time complexity are omitted.}. Shi {\it et al.}~\cite{19} improved this bound and developed an algorithm of running time $O(3^k nm)$. For the Maximum Agreement Forest problem on multiple unrooted binary phylogenetic trees, Shi {\it et al.}~\cite{19} presented the first parameterized algorithm of running time $O(4^k nm)$. To our best knowledge, there is no known parameterized algorithm for the Maximum Agreement Forest problem on multiple rooted (unrooted) multifurcating phylogenetic trees.

\smallskip

{\bf Our Contributions}. In this paper, we are focused on the fixed-parameter algorithms for the two versions (the hard version and the soft version) of the Maximum Agreement Forest problem on multiple rooted multifurcating phylogenetic trees (the {\sc Maf} problem). The general idea of our algorithms is similar to that of the previous parameterized algorithms for the Maximum Agreement Forest problem: remove edges from trees to reconcile the structural differences among them, then using the relation between the number of edges removed by the algorithm and the order of the resulting forest to design a branch-and-bound parameterized algorithm.

All previous parameterized algorithms employed the following strategy: (1) fix a tree and try to find a local structure in other trees that conflicts with the fixed tree; then (2) remove edges from the fixed tree to reconcile the structural difference. As a consequence, all branching operations are applied only on the fixed tree. Obviously, this way is convenient for analyzing the time complexity of the algorithm, because each branching operation would increase the order of the resulting forest in the fixed tree and the order of the resulting forest cannot be greater than the order of the MAF that we are looking for. However, this way does not take full advantage of the structural information given by all the trees. For example, there may exist a local structure in the fixed tree such that the corresponding branching operation on the other trees has better performance.

By careful and detailed analysis on the structures of phylogenetic trees, we propose a new branch-and-bound strategy such that the branching operations can be applied on different phylogenetic trees in the input instance. Then by making full use of special relations among leaves in phylogenetic trees, two parameterized algorithms for the {\sc Maf} problem are presented: one is for the hard version of the {\sc Maf} problem with running time $O(2.42^k m^3 n^4)$, which is the first fixed-parameter algorithm for the hard version of the problem; and the other is for the soft version of the {\sc Maf} problem with running time $O(2.74^k m^3 n^5)$, which is also the first fixed-parameter algorithm for the soft version of the problem.

The rest of the paper is structured as follows. Section 2 gives related definitions for
multifurcating phylogenetic trees and the problem formulation. Detailed presentation and
analysis of our algorithm for the hard version of the {\sc Maf} problem is given in Sections 3-5. The analysis of the algorithm for the soft version of the {\sc Maf} problem is given in Section 6, in a similar way to that for the hard version. The conclusion is presented in Section 7.

\section{Definitions and Problem Formulations}

The notations and definitions in this paper follow the ones in~\cite{19}. All graphs in our discussion are undirected. For a vertex $v$, denote the set of neighbors of $v$ by $N(v)$, and the {\it degree} of $v$ is equal to $|N(v)|$. Denote by $[u, v]$ the edge whose two ends are the vertices $u$ and $v$. A tree $T$ is a {\it single-vertex tree} if it consists of a single vertex, which is the leaf of $T$. A tree $T$ is a {\it single-edge tree} if it consists of an edge with two leaves. A tree is {\it multifurcating} if either it is a single-vertex tree or each of its vertices has degree either 1 or not less than 3. For a multifurcating tree $T$ that is not a single-vertex tree, the degree-$1$ vertices are {\it leaves} and the other vertices are {\it non-leaves}.

\subsection{$X$-tree, $X$-forest}

A {\it label-set} is a set of elements that are called ``labels''. For a label-set $X$, a multifurcating {\it phylogenetic $X$-tree} is a multifurcating tree whose leaves are labeled bijectively by the label-set $X$. A multifurcating phylogenetic $X$-tree is {\it rooted} if a particular leaf is designated as the root (so it is {\it both} a root and a leaf) -- in this case a unique ancestor-descendant relation is defined in the tree. The root of a rooted multifurcating phylogenetic $X$-tree will always be labeled by a special label $\rho$, which is always assumed to be in the label-set $X$. In the following, a rooted multifurcating phylogenetic $X$-tree is simply called an {\it $X$-tree}. As there is a bijection between the leaves of an $X$-forest and the labels in the label-set $X$, we will use, without confusion, a label in $X$ to refer to the corresponding leaf in the $X$-forest, or vice versa.

A {\it subforest} of an $X$-tree $T$ is a subgraph of $T$, and a {\it subtree} of $T$ is a connected subgraph of $T$, in both case, we assume that the subgraph contains at least one leaf in $T$. For a subtree $T'$ of a rooted $X$-tree $T$, in order to preserve the ancestor-descendant relation in $T$, a vertex in $T'$ should be defined to be the root of  $T'$. If $T'$ contains the label $\rho$, then it is the root of $T'$; otherwise, the node in $T'$ that is in $T$ the least common ancestor of all the labeled leaves in $T'$ is defined to be the root of $T'$. An {\it $X$-forest $F$} is a subforest of an $X$-tree $T$ that contains a collection of subtrees whose label-sets are disjoint such that the union of the label-sets is equal to $X$. The number of connected components in an $X$-forest $F$ is called the {\it order} of $F$, denoted by $Ord(F)$.

For any vertex $v$ in an $X$-forest $F$, denote by $L(v)$ the set containing all labels that are descendants of $v$. For any subset $V'$ of vertices in $F$, denote by $L(V')$ the union of $L(v)$ for all $v \in V'$.  For a connected component $C$ in $F$, denote by $L(C)$ the set containing all labels in $C$. For a subset $S$ of label-set $X$, where the labels in $S$ are in the same connected component of $F$, denote by $T_{F}[S]$ the minimum subtree induced by the labels of $S$ in $F$.

A subtree $T'$ of an $X$-tree may contain unlabeled vertices of degree less than $3$. In this case the {\it forced contraction} operation is applied on $T'$, which replaces each degree-$2$ vertex $v$ and its incident edges with a single edge connecting the two neighbors of $v$, and removes each unlabeled vertex that has degree 1. However, in order to preserve the ancestor-descendant relation in $T'$, if the root $r$ of $T'$ is of degree-$2$, then the operation will {\it not} be applied on $r$.
Since each connected component of an $X$-forest contains at least one labeled leaf, the forced contraction does not change the order of the $X$-forest. It is well-known (see,
e.g., \cite{10,Hallett}) that the forced contraction operation does not affect the construction of an MAF for $X$-trees. Therefore, we assume that the forced contraction is applied immediately whenever it is applicable. An $X$-forest $F$ is {\it irreducible} if the forced contraction cannot be applied to $F$. Thus, the $X$-forests in our discussion are assumed to be irreducible. With this assumption, in each (irreducible) $X$-forest $F$, the root of each connected component $T'$ is either an unlabeled vertex of degree at least 2, or the vertex labeled with $\rho$ of degree-1, or a labeled vertex of degree-0, and each unlabeled vertex in $T'$ that is not the root of $T'$ has degree at least 3.

For two $X$-forests $F_1$ and $F_2$, if there is a graph isomorphism between $F_1$ and $F_2$ in which each leaf of $F_1$ is mapped to a leaf of $F_2$ with the same label, then $F_1$ and $F_2$ are {\it isomorphic}. We will simply say that an $X$-forest $F'$ is a subforest of another $X$-forest $F$ if $F'$ is isomorphic to a subforest of $F$ (up to the forced contraction).

\subsection{Binary Resolution of $X$-forest}

An $X$-tree is {\it binary} if either it is a single-vertex tree or each of its vertices has degree either 1 or 3 (we treat the binary $X$-tree as a special type of $X$-tree). A {\it binary $X$-forest} is defined analogously.

Given two $X$-forests $F$ and $F'$, $F'$ is a {\it binary resolution} of $F$ if $F'$ is a binary $X$-forest and $F$ can be obtained by contracting some internal edges (i.e., edges between non-leaves) in $F'$. Note that if $X$-forest $F$ is binary, then itself is the unique binary resolution of $F$. Given two $X$-forests $F$ and $F'$, $F'$ is a {\it binary subforest} of $F$ if $F'$ is a binary $X$-forest, and there exists a binary resolution $F^B$ of $F$ such that $F'$ is a subforest of $F^B$.

\subsection{Agreement Forest}

Given a collection $\{F_1,F_2,\ldots,F_m\}$ of $X$-forests. An $X$-forest $F$ is a {\it hard agreement forest} for $\{F_1,F_2,\ldots,F_m\}$ if $F$ is a subforest of $F_i$, for all $1 \leq i \leq m$. An $X$-forest $F$ is a {\it soft agreement forest} for $\{F_1,F_2,\ldots,F_m\}$ if $F$ is a binary subforest of $F_i$, for all $1 \leq i \leq m$.

A {\it hard maximum agreement forest} (hMAF) for $\{F_1, F_2, \ldots, F_m\}$ is an hard agreement forest for $\{F_1, F_2, \ldots, F_m\}$ with the minimum order over all hard agreement forests for $\{F_1, F_2, \ldots, F_m\}$. The  {\it soft maximum agreement forest} (sMAF) is defined analogously.

The two versions of the Maximum Agreement Forest problem on multiple $X$-forests studied in this paper are formally defined as follows.
\begin{quote}
{\bf Hard Maximum Agreement Forest problem} ({\sc hMaf})

{\sc Input}: A set $\{F_1, \ldots, F_m\}$ of $X$-forests, and a parameter $k$\\
{\sc Output}: a hard agreement forest for $\{F_1, \ldots, F_m\}$ whose order is not larger than \\
\hspace*{16mm} $Ord(F_h)+k$, where $F_h$ is the $X$-forest in $\{F_1, \ldots, F_m\}$ that has the\\
\hspace*{16mm} largest order; or report that no such a hard agreement forest exists.\\
\end{quote}

\begin{quote}
{\bf Soft Maximum Agreement Forest problem} ({\sc sMaf})

{\sc Input}: A set $\{F_1, \ldots, F_m\}$ of $X$-forests, and a parameter $k$\\
{\sc Output}: a soft agreement forest for $\{F_1, \ldots, F_m\}$ whose order is not larger than\\
\hspace*{16mm} $Ord(F_h)+k$ , where $F_h$ is the $X$-forest in $\{F_1, \ldots, F_m\}$ that has the \\
\hspace*{16mm} largest order; or report that no such a soft agreement forest exists.
\end{quote}


\subsection{Siblings, Sibling-set, Sibling-pair}

Two leaves of an $X$-forest $F$ are {\it siblings} if they have a common parent. A {\it sibling-set} of $F$ is a set of leaves that are all siblings. A {\it maximal sibling-set} (MSS) $S$ of $F$ is a sibling-set such that the common parent $p$ of the leaves in $S$ has degree either $|S|$ if $p$ has no parent or $|S|+1$ if $p$ has a parent. A {\it sibling-pair} is an MSS that contains exact two leaves.

\subsection{Label-set Isomorphism Property, Essential Edge-set}

Two $X$-forests $F$ and $F'$ satisfy the {\it label-set isomorphism} property if for each connected component $C$ in $F$, there is a connected component $C'$ in $F'$ such that $L(C) = L(C')$. An instance of the {\sc hMaf} (or {\sc sMaf}) problem satisfies the {\it label-set isomorphism} property if any two $X$-forests in the instance satisfy the label-set isomorphism property.

Given an $X$-forest $F$ and a subset $E'$ of edges in $F$, denote by $F \setminus E'$ the $X$-forest $F$ with the edges in $E'$ removed (up to the forced contraction). The edge-set $E'$ is an {\it essential edge-set} (ee-set) of $F$ if $Ord(F \setminus E') = Ord(F) + |E'|$. Note that it is easy to test if an edge-set is an ee-set of the given $X$-forest.

\section{Instance Satisfying Label-set Isomorphism Property}

The hMAF (or sMAF) for the $X$-forests in an instance $(F_1,F_2,\ldots,F_m;k)$ of the {\sc hMaf} problem (or the {\sc sMaf} problem), is simply called the MAF for the $X$-forests in $(F_1,F_2,\ldots,F_m;k)$.
This section and the following Sections 4-5 are for the {\sc hMaf} problem.

Every MAF $F^*$ for the $X$-forests in an instance $(F_1,F_2,\ldots,F_m;k)$ of the {\sc hMaf} problem corresponds to a unique minimum subgraph $F_i^{F^*}$ of $F_i$, for $1 \leq i \leq m$, which consists of the paths in $F_i$ that connect the leaves in the same connected component in $F^*$. Thus, for any edge $e$ in $F_i$, without any confusion, we can simply say that $e$ is in or is not in the MAF $F^*$, as long as $e$ is in or is not in the corresponding subgraph $F_i^{F^*}$, respectively.

Given an instance $(F_1,F_2,\ldots,F_m;k)$ of the {\sc hMaf} problem. If $(F_1,F_2,\ldots,F_m;k)$ does not satisfy the label-set isomorphism property, then two rules given in the following subsection can be applied to eliminate the difference among the label-sets of the connected components in the $X$-forests in $(F_1,F_2,\ldots,F_m;k)$. Denote by $Ord_{max}(F_1,F_2,\ldots,F_m;k)$ the maximum order of an  $X$-forest in $(F_1,F_2,\ldots,F_m;k)$.

\subsection{Two Rules}

\noindent{\bf Reduction Rule 1.} Let $\mathcal{C}_{F_i} = \{C_1,\ldots,C_t\}$ ($t \geq 1$) be a subset of the connected components in the $X$-forest $F_i$, $1 \leq i \leq m$. If there is a vertex $v$ in a connected component $C$ of the $X$-forest $F_j$, $j \neq i$, such that $L(v) = L(C) \cap (L(C_1) \cup \ldots \cup L(C_t))$, then remove the edge $e$ between $v$ and $v$'s parent (if one exists) in $F_j$.

\smallskip

For the situation of Reduction Rule 1, we say that Reduction Rule 1 is {\it applicable on $F_j$ relative to $F_i$}.
Let $(F_1,\ldots,F_j \setminus \{e\},\ldots,F_m;k')$ be the instance obtained by applying Reduction Rule 1 on $(F_1,F_2,\ldots,F_m;k)$ with edge $e$ removed from $F_j$. By the formulation of the {\sc hMaf} problem given in the previous section, we have that $Ord_{max}(F_1,F_2,\ldots,F_m;k) + k = Ord_{max}(F_1,\ldots,F_j \setminus \{e\},\ldots,F_m;k') + k'$. Thus, if $Ord(F_j) = Ord_{max}(F_1,F_2,\ldots,F_m;k)$, then $Ord_{max}(F_1,\ldots,F_j \setminus \{e\},\ldots,F_m;k') = Ord(F_j \setminus \{e\}) = Ord_{max}(F_1,F_2,\ldots,F_m;k)+1$ and $k' = k-1$, otherwise, $k' = k$. For instances $(F_1,F_2,\ldots,F_m;k)$ and $(F_1,\ldots,F_j \setminus \{e\},\ldots,F_m;k')$, we have the following lemma.

\begin{lemma}
\label{lem:Reduction Rule 1}
Instances $(F_1,F_2,\ldots,F_m;k)$ and $(F_1,\ldots,F_j \setminus \{e\},\ldots,F_m;k')$ have the same collection of solutions.

\begin{proof}
Firstly, we show that every agreement forest for $\{F_1,F_2,\ldots,F_m\}$ is also an agreement forest for $\{F_1,\ldots,F_j \setminus \{e\},\ldots,F_m\}$.
Suppose $F^*$ is an agreement forest for $\{F_1,F_2,\ldots,F_m\}$. Let $Y= L(C_1) \cup \ldots \cup L(C_t)$ and $Y'= X \setminus Y$. Since $F^*$ is a subforest of $F_i$, for each connected component $C_s$ in $\mathcal{C}_{F_i}$, $1 \leq s \leq t$, we have that any label of $L(C_s)$ cannot be in the same connected component with any label of $X \setminus L(C_s)$ in $F^*$. Thus, any label of $Y$ cannot be in the same connected component with any label of $Y'$ in $F^*$.

Suppose that edge $e$ is in $F^*$. Then there would exist a path in $F^*$ that connects a label of $Y$ and a label of $Y'$, contradicting the fact that any label of $Y$ cannot be in the same connected component with any label of $Y'$ in $F^*$. Thus, edge $e$ cannot be in $F^*$ and $F^*$ is still a subforest of $F_j \setminus \{e\}$. Therefore, $F^*$ is also an agreement forest for $\{F_1,\ldots,F_j \setminus \{e\},\ldots,F_m\}$.

In the following, we show that every agreement forest for $\{F_1,\ldots,F_j \setminus \{e\},\ldots,F_m\}$ is also an agreement forest for $\{F_1,F_2,\ldots,F_m\}$. Suppose that $F^*$ is an agreement forest for $\{F_1,\ldots,F_j \setminus \{e\},\ldots,F_m\}$. Since $F^*$ is a subforest of $F_j \setminus \{e\}$, $F^*$ is also a subforest of $F_j$. Therefore, $F^*$ is also an agreement forest for $\{F_1,F_2,\ldots,F_m\}$.

By above analysis, $\{F_1,F_2,\ldots,F_m\}$ and $\{F_1,\ldots,F_j \setminus \{e\},\ldots,F_m\}$ have the same collection of agreement forests. Since $Ord_{max}(F_1,F_2,\ldots,F_m;k) + k = Ord_{max}(F_1,\ldots,F_j \setminus \{e\},\ldots,F_m;k') + k'$, $X$-forest $F$ is a solution of $(F_1,F_2,\ldots,F_m;k)$ if and only if $F$ is also a solution of $(F_1,\ldots,F_j \setminus \{e\},\ldots,F_m;k')$.
\end{proof}
\end{lemma}

In the following discussion, we assume that Reduction Rule 1 is not applicable on the
given instances. Our second rule is a branching rule. We first give some related definitions.
We say that a branching rule is {\it safe} if on an instance $(F_1, \ldots, F_m; k)$ it
produces a collection $S$ of instances such that $(F_1, \ldots, F_m; k)$ is a yes-instance
if and only if at least one of the instances in $S$ is a yes-instance. A branching rule
satisfies the recurrence relation $T(k) = T(k_1)+ \ldots + T(k_r)$ if on an instance
$(F_1, \ldots, F_m; k)$, it produces $r$ instances $(F^1_1, \ldots, F^1_m; k_1)$, $\ldots$,
$(F^r_1, \ldots, F^r_m; k_r)$. We also say that the branching rule satisfies the recurrence
relation $T(k) \leq T(k'_1)+ \ldots + T(k'_t)$ ($t \geq 2$) if the positive root of the
characteristic polynomial of $T(k) = T(k_1)+ \ldots + T(k_r)$ is not larger than that of
$T(k) = T(k'_1)+ \ldots + T(k'_t)$ (see \cite{vcold} for more discussions). Moreover, we
assume that the function $T(k)$ is non-decreasing.

\smallskip

\noindent {\bf Case 1.} For a connected component $C$ in $F_i$, $1 \leq i \leq m$, there exists a vertex $v$ with two children $c_1$ and $c_2$ in the connected component $C'$ of $F_j$, $j \neq i$, such that $L(c_1) \subseteq L(C)$ and $L(c_2) \cap L(C) = \emptyset$.

\smallskip

\noindent {\bf Branching Rule 1.} Branch into two ways: [1] remove the edge $[v,c_1]$ in $F_j$; [2] remove the edge $[v,c_2]$ in $F_j$.

\smallskip

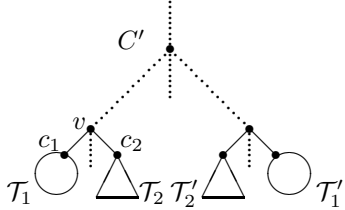
\begin{figure}[h]
\begin{center}
\begin{picture}(300,40)
\put(100,-35){\begin{picture}(0,0)

 \put(50,70){\circle*{3}}
 \multiput(50,70)(0,2){10}{\circle*{1}}
 \multiput(50,70)(-2,-2){15}{\circle*{1}}
 \multiput(50,70)(2,-2){15}{\circle*{1}}
 \multiput(50,70)(0,-2){10}{\circle*{1}}

 \multiput(20,40)(0,-2){8}{\circle*{1}}
 \multiput(80,40)(0,-2){8}{\circle*{1}}
 \put(20,40){\circle*{3}}
 \put(80,40){\circle*{3}}

 \put(13,40){\small $v$}

 \put(20,40){\line(-1,-1){10}}
 \put(10,30){\circle*{3}}
 \put(0,33){\small $c_1$}
 \put(7,23){\circle{16}}

 \put(20,40){\line(1,-1){10}}
 \put(30,30){\circle*{3}}
 \put(31,33){\small $c_2$}
 \put(30,30){\line(-1,-2){8}}
 \put(30,30){\line(1,-2){8}}
 \put(22,14){\line(1,0){16}}

 \put(80,40){\line(1,-1){10}}

 \put(80,40){\line(-1,-1){10}}
 \put(70,30){\circle*{3}}
 \put(70,30){\line(-1,-2){8}}
 \put(70,30){\line(1,-2){8}}
 \put(62,14){\line(1,0){16}}

 \put(90,30){\circle*{3}}
 \put(95,23){\circle{16}}

 \put(38,13){\small $\mathcal{T}_2$}
 \put(-12,13){\small $\mathcal{T}_1$}
 \put(50,13){\small $\mathcal{T}'_2$}
 \put(105,13){\small $\mathcal{T}'_1$}

 \put(30,70){\small $C'$}
\end{picture}}

\end{picture}

\end{center}
\vspace*{5mm}
\caption{The general structure of connected component $C'$. The triangles and circles denote subtrees. The label-sets of $\mathcal{T}_1$ and $\mathcal{T}'_1$ belong to $L(C)$, where $L(\mathcal{T}_1) = L(c_1)$. The label-sets of $\mathcal{T}_2$ and $\mathcal{T}'_2$ do not belong to $L(C)$, where $L(\mathcal{T}_2) = L(c_2)$. }
\label{fig:branching Rule 1}
\end{figure}

Figure~\ref{fig:branching Rule 1} gives an illustration of Case 1, for which we will
say that Branching Rule 1 is {\it applicable on $F_j$ relative to $F_i$}. It is
necessary to remark that there exists at least one label in $L(C) \setminus L(c_1)$
that is in the connected component $C'$ -- otherwise, the edge $[v,c_1]$ could be
removed by Reduction Rule 1. We have the following two observations for Case 1.

\begin{observation}
\label{obs:1}
For each of the two edges $[v, c_1]$ and $[v, c_2]$, there are two labels such that
the edge is on the path connecting the two labels in $F_j$, and the two labels are in
the same connected component of $F_i$.
\end{observation}

\begin{observation}
\label{obs:2}
For any $X$-forest $F_s$ in $(F_1, \ldots, F_m; k)$, $s \neq j$, there are two labels
$l_1 \in L(c_1)$ and $l'_1 \in L(C') \setminus L(c_1)$ that are in the same connected
component of $F_s$. There are also two labels $l_2\in L(c_2)$ and $l'_2\in L(C')\setminus L(c_2)$
that are in the same connected component of $F_s$.
\end{observation}

\begin{lemma}
\label{lem:Branching Rule 1}
Branching Rule 1 is safe.

\begin{proof}
Let $F^*$ be an agreement forest for the $X$-forests in $(F_1, \ldots, F_m; k)$. If both edges
$[v,c_1]$ and $[v,c_2]$ are in $F^*$, then there would be a label in $L(C)$ and a label in
$X \setminus L(C)$ that are in the same connected component of $F^*$. However, this is impossible
because $C$ is a connected component of $F_i$ and $F^*$ is a subforest of $F_i$, so a connected
component of $F^*$ cannot have both labels in $L(C)$ and labels in $X \setminus L(C)$.

Thus, at least one of the edges $[v,c_1]$ and $[v,c_2]$ is not in $F^*$, which is an arbitrary
agreement forest for $(F_1, \ldots, F_m; k)$. Consequently, at least one of the two branches in
Branching Rule 1 is correct. Thus, the rule is safe.
\end{proof}
\end{lemma}

\begin{lemma}
\label{lem:label-isomorphism}
Any instance of the {\sc hMaf} problem on which Reduction Rule 1 and Branching Rule 1 are unapplicable, satisfies the label-set isomorphism property.

\begin{proof}
It suffices to prove that if neither of Reduction Rule 1 and Branching Rule 1 is
applicable on any one of the two $X$-forests $F$ and $F'$ relative to the other,
then $F$ and $F'$ satisfy the label-set isomorphism property. Suppose
for the contrary that there are two connected components $C$ and $C'$ of $F$ and $F'$
respectively, such that $L(C) \neq L(C')$ and $L(C) \cap L(C') \neq \emptyset$.

(1). Suppose that one of $L(C)$ and $L(C')$ is a proper subset of the other. Because of
the symmetry, we can assume $L(C) \subsetneqq L(C')$. If there is a $v$ in $C'$ such
that $L(v) = L(C)$, then the edge between $v$ and the parent of $v$ would be removed by
Reduction Rule 1, contradicting the assumption that Reduction Rule 1 is not applicable
on $F'$ relative to $F$. If there is no such a vertex $v$, then there must be a vertex
$v'$ with two children $v_1$ and $v_2$ in $C'$ such that $L(v_1) \subsetneqq L(C)$ and
$L(v_2) \cap L(C) = \emptyset$. Then, the edges $[v',v_1]$ and $[v',v_2]$ would be removed
by Branching Rule 1, contradicting the assumption that Branching Rule 1 is not applicable
on $F'$ relative to $F$.

(2). If neither of $L(C)$ and $L(C')$ is a proper subset of the other, then there is
a vertex $v$ with two children $v_1$ and $v_2$ in $C$ such that $L(v_1) \varsubsetneqq L(C')$
and $L(v_2) \cap L(C') = \emptyset$. If $L(v_1) = L(C) \cap L(C')$, then the edge $[v, v_1]$
would be removed by Reduction Rule 1; if $L(v_2) = L(C) \setminus L(C')$, then the edge
$[v, v_2]$ would be removed by Reduction Rule 1. If neither of these is the case, then the
edges $[v,v_1]$ and $[v ,v_2]$ would be removed by Branching Rule 1. Thus, all cases would
contradict the assumption of the lemma.

Summarizing the above discussions gives the proof of the lemma.
\end{proof}
\end{lemma}

Let $(F_1, \ldots, F_m; k)$ be an arbitrary instance of the {\sc hMaf} problem on which
Reduction Rule 1 is not applicable. If $(F_1, \ldots, F_m; k)$ does not satisfy the
label-set isomorphism property, then by Lemma~\ref{lem:label-isomorphism}, Branching
Rule 1 can be applied, resulting in two instances. If the resulting instances do not
satisfy the label-set isomorphism property, then we can recursively apply Reduction
Rule 1 and Branching Rule 1, repeatedly, until all instances constructed in this process
satisfy the label-set isomorphism property.

Let $(F'_1, \ldots, F'_m; k')$ be any of these constructed instances. It is critical
for us to know how many times Branching Rule 1 is applied in the process from
$(F_1, \ldots, F_m; k)$ to $(F'_1, \ldots, F'_m; k')$. To answer this is not easy
because Branching Rule 1 can remove edges from different $X$-forests in the instance.
In the following, we first analyze a special process for two $X$-forests $F_p$ and
$F_q$ ($p < q$) in the instance $(F_1, \ldots, F_m; k)$,
which is called the {\it 2-BR-process} on $F_p$ and $F_q$. Note that Reduction Rule 1
is assumed not applicable on $(F_1, \ldots, F_m; k)$. The 2-BR-process on $F_p$ and
$F_q$ consists of the following three stages. Initialize the collection $\mathcal{C}$
with $\{(F_1, \ldots, F_m; k)\}$.

\smallskip

\noindent{\bf Stage-1.} For an instance $(F'_1, \ldots,F'_m; k')$ in $\mathcal{C}$,
if Branching Rule 1 is applicable on $F'_p$ relative to $F'_q$ (or on $F'_q$ relative
to $F'_p$), then apply Branching Rule 1 on $F'_p$ and $F'_q$, and replace the
instance $(F'_1, \ldots,F'_m; k')$ in $\mathcal{C}$ with the two instances resulted
from the application of the rule.

\smallskip

\noindent{\bf Stage-2.} For an instance $(F'_1, \ldots,F'_m; k')$ in $\mathcal{C}$,
if Reduction Rule 1 is applicable, then repeatedly apply Reduction Rule 1 until
the rule is not applicable. Replace the instance $(F'_1, \ldots,F'_m; k')$ in $\mathcal{C}$
with the resulting instance.

\smallskip

\noindent{\bf Stage-3.} Repeatedly apply Stage-1 and Stage-2, in this order, on any  instance
$(F'_1, \ldots,F'_m; k')$ in $\mathcal{C}$ in which $F'_p$ and $F'_q$ do not satisfy the
label-set isomorphism property.

\smallskip

At the end of this process on $F_p$ and $F_q$, in every instance $(F'_1, \ldots, F'_m; k')$
in $\mathcal{C}$, the $X$-forests $F'_p$ and $F'_q$ satisfy the label-set isomorphism property.

\subsection{2-BR-process on $F_p$ and $F_q$}

Let $(F_1^*, \ldots, F_m^*; k^*)$ be an instance obtained by the 2-BR-process on
$F_p$ and $F_q$ of $(F_1, \ldots, F_m; k)$, and let $S = \{e_1, \ldots, e_h\}$
($h \geq 1$) be the sequence of edges removed by Reduction Rule 1 and Branching
Rule 1 during the 2-BR-process on $F_p$ and $F_q$ from $(F_1, \ldots, F_m; k)$ to
$(F_1^*, \ldots, F_m^*; k^*)$, in which $e_l$, $1 \leq l \leq h$, is an edge of
the instance $I_l = (F_1^l, \ldots, F_m^l; k^l)$. Let $S_B$ be the subsequence of
$S$ that contains all edges removed by Branching Rule 1.

Since $F_p^*$ and $F_q^*$ satisfy the label-set isomorphism property, $Ord(F_p^*)=Ord(F_q^*)$.
We study the relations among $Ord(F_p)$, $Ord(F_q)$, $Ord(F_p^*)$, and $|S_{B}|$. By
Observation~\ref{obs:1}, for an edge $e_r$ in $S_B$, there is a label-pair $(a,a')$ such
that $e_r$ is on the unique path connecting $a$ and $a'$ in $F_p^r$ (or $F_q^r$), and $a$
and $a'$ are also in the same connected component of $F_q^r$ (or $F_p^r$). We call the
label-pair $(a,a')$ a {\it connected label-pair} for the edge $e_r$.

We explain how to find a connected label-pair for the edge $e_r$. Without loss of generality,
assume that $e_r = [v,v']$ is in $F_p^r$, where $v$ is the parent of $v'$. For each connected
component $C'$ of $F_q^r$, check if $L(C') \cap L(v') \neq \emptyset$ and
$L(C') \cap (L(C) \setminus L(v')) \neq \emptyset$, where $C$ is the connected component of
$F_p^r$ containing the edge $e_r$. Note that there must be a connected component $C'$
in $F_q^r$ that satisfies these conditions  -- otherwise, Reduction Rule 1 would be applicable
on the vertex $v'$ in $F_p^r$ relative to $F_q^r$. Now arbitrarily picking two labels from
$L(C') \cap L(v')$ and $L(C') \cap (L(C) \setminus L(v'))$, respectively, will give a connected
label-pair for $e_r$.

Let $S_{lp}$ be the sequence that contains a connected label-pair for each edge in $S_B$. Since
there may be a label that appears in more than one connected label-pair in $S_{lp}$, for the
simplicity of analysis, we construct several {\it dummy} labels for it. For example, if label $x$
appears in three connected label-pairs in $S_{lp}$, then we construct three dummy labels $x_1$,
$x_2$, and $x_3$ for it, and replace the label $x$ in the three connected label-pairs with $x_1$,
$x_2$, and $x_3$, respectively. By this operation, each (dummy) label appears in only one
connected label-pair in $S_{lp}$. We also say that $x$ is the dummy label of itself if label $x$
appears in only one connected label-pair in $S_{lp}$.

Let $X(S_{lp})$ be the set of the dummy labels that appear in $S_{lp}$. We say that the label
$x \in X$ is in $X(S_{lp})$ if some dummy label of $x$ is in $X(S_{lp})$, that the connected
component $C$ of an $X$-forest contains the dummy label $x \in X(S_{lp})$ if $x$ is a dummy
label of some label in $C$, and that two dummy labels $x_1$ and $y_1$ of $X(S_{lp})$ are not
in the same connected component of an $X$-forest $F$ if labels $x$ and $y$ are not in the same
connected component of $F$, where $x_1$ and $y_1$ are the dummy labels of $x$ and $y$,
respectively.

\begin{lemma}
\label{lem:equal}
$Ord(F_p^*) = Ord(F_q^*) \geq |S_{B}| + \max\{Ord(F_p), Ord(F_q)\}$.

\begin{proof}
Given a connected component $C$ of an $X$-forest $F$, denote by $X(C)$ the subset of
$X(S_{lp})$ such that for each label $x \in L(C)$, if $x$ is in $X(S_{lp})$, then $X(C)$
contains all dummy labels of $x$ that are in $X(S_{lp})$ -- otherwise, $X(C)$ does not
contain any dummy label of $x$. Note that all dummy labels of a label are always in the
same connected component of $F$, hence $X(C)$ either contains all dummy labels of the
label or contains none.

Let $C_1$, $\ldots$, $C_t$ be the connected components of $F^{*}_p$. If $|X(C_s)| \leq 1$
for all $1 \leq s \leq t$, then the lemma obviously holds true, since the two dummy labels of
each connected label-pair are in the same connected component of $F_p$. Thus, we can
assume that there is a connected component $C_z$ of $F^{*}_p$ such that $|X(C_z)| \geq 2$.

By symmetry, we can assume $Ord(F_p) \geq Ord(F_q)$, so the lemma claims
$Ord(F_p^*) \geq |S_{B}| + Ord(F_p)$. For an $X$-forest $F$, denote by $N_{F,{X(S_{lp})}}$
the number of connected components of $F$ that contain dummy labels in $X(S_{lp})$,
and by $N_{F,{\overline{X(S_{lp})}}}$ the number of connected components of $F$ that do
not contain any dummy label in $X(S_{lp})$. Then
$Ord(F_p) = N_{F_p,{X(S_{lp})}} + N_{F_p,{\overline{X(S_{lp})}}}$,
$Ord(F^{*}_p) = N_{F^{*}_p,{X(S_{lp})}} + N_{F^{*}_p,{\overline{X(S_{lp})}}}$, and
the lemma can be proved by showing
\begin{equation}
\label{inequality1}
N_{F^{*}_p,X(S_{lp})} + N_{F^{*}_p,{\overline{X(S_{lp})}}} \geq
     |S_{B}| + N_{F_p,{X(S_{lp})}} + N_{F_p,{\overline{X(S_{lp})}}}.
\end{equation}

We prove the inequality (\ref{inequality1}) by induction on $|S_{B}|$. If $|S_{B}|=0$, then
$S_{lp} = \emptyset$ and $N_{F^{*}_p,X(S_{lp})} = N_{F_p,{X(S_{lp})}} = 0$. Since
$Ord(F^{*}_p) = N_{F^{*}_p,{\overline{X(S_{lp})}}} \geq Ord(F_p) = N_{F_p,{\overline{X(S_{lp})}}}$,
the inequality (\ref{inequality1}) holds true for $|S_{B}|=0$.

Now consider $|S_{B}|=1$ and $S_{lp} = \{(a,a')\}$. Since $a$ and $a'$ are in the same
connected component of $F_p$ and are in different connected components of $F^{*}_p$,
$N_{F_p,{X(S_{lp})}} = 1$ and $N_{F^{*}_p,X(S_{lp})} = 2$. Combining this with the fact
$N_{F^{*}_p,{\overline{X(S_{lp})}}} \geq N_{F_p,{\overline{X(S_{lp})}}}$ gives the inequality
(\ref{inequality1}) when $|S_{B}|=1$.

For the general case $|S_{B}|=r+1$, where $r \geq 1$, let $S_B = \{e_{i_1},\ldots\, e_{i_{r}},e_{i_{r+1}}\}$,
and let $(a,a')$ be the connected label-pair for $e_{i_1}$. Let $S'_{lp} = S_{lp} \setminus \{(a,a')\}$
and $S'_B = S_B \setminus \{e_{i_1}\}$. Since $i_s + 1 \leq i_{s +1}$ for any $1 \leq s \leq r$,
we have
\begin{eqnarray*}
  N_{F^{i_s+1}_p,{X(S_{lp})}} \leq  N_{F^{i_{s+1}}_p,{X(S_{lp})}}, \ \ \
  N_{F^{i_s+1}_p,{\overline{X(S_{lp})}}} \leq N_{F^{i_{s+1}}_p,{\overline{X(S_{lp})}}}, \\
  N_{F^{i_s+1}_p,{X(S'_{lp})}} \leq  N_{F^{i_{s+1}}_i,{X(S'_{lp})}}, \ \ \
  N_{F^{i_s+1}_p,{\overline{X(S'_{lp})}}} \leq N_{F^{i_{s+1}}_p,{\overline{X(S'_{lp})}}}.
\end{eqnarray*}
Since we assumed that Reduction Rule 1 is not applicable on $(F_1, \ldots, F_m; k)$, the
first edge $e_{i_1}$ in $S_B$ is also the first edge in $S$, i.e., $e_{i_1} = e_1$, and
$(F^{i_1}_1, \ldots,F^{i_1}_m; k^{i_1}) = (F^{1}_1, \ldots, F^{1}_m; k^{1}) = (F_1, \ldots, F_m; k)$.

By the inductive hypothesis for $|S_{B}|=r$, we have the following inequality for $F_{p}^{i_2}$,
$F_{q}^{i_2}$, $S'_{lp}$, and $S'_{B}$:
\begin{equation}
\label{inequality1-1}
N_{F^{*}_p,X(S'_{lp})} + N_{F^{*}_p,{\overline{X(S'_{lp})}}}  \geq |S'_{B}| + \max\{N_{F^{i_2}_p,{X(S'_{lp})}}
      + N_{F^{i_2}_p,{\overline{X(S'_{lp})}}}, N_{F^{i_2}_q,{X(S'_{lp})}}
      + N_{F^{i_2}_q,{\overline{X(S'_{lp})}}}\}.
\end{equation}
We divide into two cases
$N_{F^{i_2}_p,{X(S'_{lp})}} + N_{F^{i_2}_p,{\overline{X(S'_{lp})}}} \geq
N_{F^{i_2}_q,{X(S'_{lp})}} + N_{F^{i_2}_q,{\overline{X(S'_{lp})}}}$ and
$N_{F^{i_2}_p,{X(S'_{lp})}} + N_{F^{i_2}_p,{\overline{X(S'_{lp})}}} <
N_{F^{i_2}_q,{X(S'_{lp})}} + N_{F^{i_2}_q,{\overline{X(S'_{lp})}}}$.

\smallskip

\noindent Case 1. $N_{F^{i_2}_p,{X(S'_{lp})}} + N_{F^{i_2}_p,{\overline{X(S'_{lp})}}}
         \geq N_{F^{i_2}_q,{X(S'_{lp})}} + N_{F^{i_2}_q,{\overline{X(S'_{lp})}}}$.
By the inequality~(\ref{inequality1-1}), we have
\[ N_{F^{*}_p,X(S'_{lp})} + N_{F^{*}_p,{\overline{X(S'_{lp})}}}
   \geq |S'_{B}| + N_{F^{i_2}_p,{X(S'_{lp})}} + N_{F^{i_2}_p,{\overline{X(S'_{lp})}}}.\]
Since $N_{F^{i_1+1}_p,{X(S'_{lp})}} \leq  N_{F^{i_{2}}_p,{X(S'_{lp})}}$ and
$N_{F^{i_1+1}_p,{\overline{X(S'_{lp})}}} \leq N_{F^{i_{2}}_p,{\overline{X(S'_{lp})}}}$,
we have
\begin{equation}
\label{inequality2}
N_{F^{*}_p,X(S'_{lp})} + N_{F^{*}_p,{\overline{X(S'_{lp})}}} \geq
 |S'_{B}| + N_{F^{i_1+1}_p,{X(S'_{lp})}} + N_{F^{i_1+1}_p,{\overline{X(S'_{lp})}}}.
\end{equation}

Let $C_a$ and $C_{a'}$ be the connected components of $F_p^{i_1+1}$ that contain
$a$ and $a'$, respectively. We divide Case 1 into three subcases:

Case 1.1.\ $|X(C_{a})| > 1$ and $|X(C_{a'})| > 1$. In this case, we have
\[ N_{F^{i_1+1}_p,{X(S'_{lp})}} \geq N_{F^{i_1}_p,{X(S'_{lp})}} + 1
   = N_{F^{i_1}_p,{X(S_{lp})}} + 1 = N_{F_p,{X(S_{lp})}} + 1 \]
and
\[ N_{F^{i_1+1}_p,{\overline{X(S'_{lp})}}} \geq N_{F^{i_1}_p,{\overline{X(S'_{lp})}}}
   \geq N_{F^{i_1}_p,{\overline{X(S_{lp})}}} =  N_{F_p,{\overline{X(S_{lp})}}}. \]
Thus, combining with the inequality~(\ref{inequality2}) and $|S_B| = |S'_B|+1$, we get
\begin{equation}
\label{cheneq1}
   N_{F^{*}_p,X(S_{lp})} + N_{F^{*}_p,{\overline{X(S_{lp})}}}  = 
   N_{F^{*}_p,X(S'_{lp})} + N_{F^{*}_p,{\overline{X(S'_{lp})}}} 
   \geq 
   |S_{B}| + N_{F_p,{X(S_{lp})}} + N_{F_p,{\overline{X(S_{lp})}}}.  
\end{equation}

Case 1.2.\ $|X(C_{a})| > 1$ and $|X(C_{a'})| = 1$. In this case we have
\[  N_{F^{i_1+1}_p,{X(S'_{lp})}} \geq N_{F^{i_1}_p,{X(S'_{lp})}} =
  N_{F^{i_1}_p,{X(S_{lp})}} = N_{F_p,{X(S_{lp})}} \]
and
\[ N_{F^{i_1+1}_p,{\overline{X(S'_{lp})}}} \geq N_{F^{i_1}_p,{\overline{X(S'_{lp})}}} + 1
  = N_{F^{i_1}_p,{\overline{X(S_{lp})}}} + 1 =  N_{F_p,{\overline{X(S_{lp})}}} + 1. \]
Combining with (\ref{inequality2}) and $|S_B| = |S'_B|+1$ gives the relation (\ref{cheneq1})
again.

Case 1.3.\ $|X(C_{a})| = 1$ and $|X(C_{a'})| = 1$. We have
\[ N_{F^{i_1+1}_p,{X(S'_{lp})}} =
   N_{F^{i_1}_p,{X(S'_{lp})}} = N_{F^{i_1}_p,{X(S_{lp})}} - 1 = N_{F_p,{X(S_{lp})}} -1\]
and
\[ N_{F^{i_1+1}_p,{\overline{X(S'_{lp})}}} \geq N_{F^{i_1}_p,{\overline{X(S'_{lp})}}} + 1
  = N_{F^{i_1}_p,{\overline{X(S_{lp})}}} + 2 = N_{F_p,{\overline{X(S_{lp})}}} + 2. \]
Combining with (\ref{inequality2}) and $|S_B| = |S'_B|+1$ gives the relation (\ref{cheneq1})
again.

Therefore, for Case 1, the inequality (\ref{inequality1}) always holds true.

Case 2. \ $N_{F^{i_2}_p,{X(S'_{lp})}} + N_{F^{i_2}_p,{\overline{X(S'_{lp})}}} <
   N_{F^{i_2}_q,{X(S'_{lp})}} + N_{F^{i_2}_q,{\overline{X(S'_{lp})}}}$. By the
inequality~(\ref{inequality1-1}), we have
\begin{eqnarray}
N_{F^{*}_p,X(S'_{lp})} + N_{F^{*}_p,{\overline{X(S'_{lp})}}}
  & \geq & |S'_B| + N_{F^{i_2}_q,{X(S'_{lp})}} + N_{F^{i_2}_q,{\overline{X(S'_{lp})}}} \label{cheneq2} \\
  & \geq & |S'_B| + N_{F^{i_2}_p,{X(S'_{lp})}} + N_{F^{i_2}_p,{\overline{X(S'_{lp})}}} + 1 \nonumber \\
  & = & |S_B| + N_{F^{i_2}_p,{X(S'_{lp})}} + N_{F^{i_2}_p,{\overline{X(S'_{lp})}}}. \nonumber
\end{eqnarray}
Since $Ord(F^{i_2}_p) \geq Ord(F^{i_1+1}_p) \geq Ord(F^{i_1}_p) = Ord(F_p)$, we have
\[ N_{F^{i_2}_p,{X(S'_{lp})}} + N_{F^{i_2}_p,{\overline{X(S'_{lp})}}} =
   Ord(F^{i_2}_p) \geq Ord(F_p) = N_{F_p,{X(S_{lp})}} + N_{F_p,{\overline{X(S_{lp})}}}. \]
Combining this with (\ref{cheneq2}) gives the inequality (\ref{inequality1}).

Thus, the inequality~(\ref{inequality1}) holds true, which implies the lemma.
\end{proof}
\end{lemma}

\subsection{The Extension of $2$-BR-process: $m$-BR-process}

Based on the $2$-BR-process, we present a process named {\it $m$-BR-process}. Let
$(F_1, \ldots, F_m; k)$ be an instance of the {\sc hMaf} problem, on which Reduction
Rule 1 may be applicable. The $m$-BR-process on $(F_1, \ldots, F_m; k)$ consists of
the following two stages.

\medskip

\noindent{\bf Stage-1.} Apply Reduction Rule 1 on $(F_1, \ldots, F_m; k)$ until it
becomes unapplicable. Let $\mathcal{C}$ be the collection containing the resulting instance.

\medskip

\noindent{\bf Stage-2.} While there is an instance $(F'_1, \ldots, F'_m; k')$ in $\mathcal{C}$
in which there are two $X$-forests $F'_s$ and $F'_t$ having the first and second largest orders
respectively that do not satisfy the label-set isomorphism property, apply the 2-BR-process on
$F'_s$ and $F'_t$, and update the collection $\mathcal{C}$.

\begin{corollary}
\label{cor:cor1}
For any instance $(F^*_1, \ldots, F^*_m; k^*)$ obtained by the $m$-BR-process on
$(F_1, \ldots, F_m; k)$, $Ord(F^*_1) \geq |S_B| + Ord_{\max}(F_1, \ldots, F_m; k)$, where
$S_B$ is the set of edges that are removed by Branching Rule 1 during the $m$-BR-process
from $(F_1, \ldots, F_m; k)$ to $(F^*_1, \ldots, F^*_m; k^*)$.

\begin{proof}
Suppose that the sequence of the executions of the 2-BR-process during the the $m$-BR-process
from $(F_1, \ldots, F_m; k)$ to $(F^*_1, \ldots, F^*_m; k^*)$ is $\{P_1,\ldots,P_j\}$, where
$P_i$ is the execution of the 2-BR-process that is applied on $F_{x_i}^{i}$ and $F_{y_i}^{i}$
of the instance $(F_1^i, \ldots, F_m^i; k^i)$ (assume $Ord(F_{x_i}^{i}) \geq Ord(F_{y_i}^{i})$).
Denote by $S_B^i$ the sequence of edges removed by Branching Rule 1 during the 2-BR-process
$P_i$ from $(F_1^i, \ldots, F_m^i; k^i)$ to $(F_1^{i+1}, \ldots, F_m^{i+1}; k^{i+1})$.
Apparently, $(F^1_1, \ldots,F^1_m; k^1)$ is the instance obtained by Stage-1 of the
$m$-BR-process, and $(F^*_1, \ldots, F^*_m; k^*)$ is the instance obtained by $P_j$.
For Stage-2, by Lemma~\ref{lem:equal}, for any $1 \leq i \leq j-1$, we have that
\[ Ord(F_{x_{i+1}}^{i+1}) \geq Ord(F_{x_{i}}^{i+1}) \geq |S_B^i| + Ord(F_{x_i}^{i}), \]
and for $i = j$, we have that
\[ Ord(F_1^*) \geq |S_B^j| + Ord(F_{x_{j}}^{j}). \]

By summing the inequalities for all $1 \leq i \leq j$, we get
\[ Ord(F_1^{*}) \geq |S_B| + Ord(F_{x_1}^{1}),  \]
where $S_B = \sum^{j}_{i=1}S^i_B$. Now from the fact
$Ord(F_{x_1}^{1}) \geq Ord_{\max}(F_1, \ldots, F_m; k)$, we get
$Ord(F^*_1) \geq |S_B| + Ord_{\max}(F_1, \ldots, F_m; k)$.
\end{proof}
\end{corollary}

To analyze the $m$-BR-process, we regard it as a branching process. Let
$\mathcal{C} = \{(F^{1}_1, \ldots, F^{1}_m; k^{1}), \ldots, (F^{q}_1, \ldots, F^{q}_m; k^{q})\}$
be the collection of instances obtained by the $m$-BR-process on $(F_1, \ldots, F_m; k)$, and
let $S^{p}_B$, $1 \leq p \leq q$, be the set of edges removed by Branching Rule 1 during the
$m$-BR-process from $(F_1, \ldots, F_m; k)$ to $(F^{p}_1, \ldots, F^{p}_m; k^{p})$. Let
$r = \min\{|S^{1}_B|, \ldots, |S^{q}_B|\}$ and $h = \max\{|S^{1}_B|, \ldots, |S^{q}_B|\}$,
and let $\mathcal{C}_l$ be the subset of $\mathcal{C}$ that contains all $(F^p_1, \ldots, F^p_m; k^p)$
in $\mathcal{C}$ such that $|S^p_B| = l$, for any $r \leq l \leq h$.

\begin{theorem}
\label{theo:m-BR-process}
If Branching Rule 1 is applied on $(F_1, \ldots, F_m; k)$ during the $m$-BR-process, then it
satisfies the recurrence relation $T(k) \leq 2 T(k-1)$.

\begin{proof}
For any instance $(F^{p}_1, \ldots, F^{p}_m; k^{p})$ in $\mathcal{C}$, by
Corollary~\ref{cor:cor1}, $k^p \leq k - |S^p_B|$. Thus, $T(k^p) \leq T(k - |S^p_B|)$
for all $1 \leq p \leq q$. This gives
$T(k) = T(k^1) + \ldots + T(k^q) \leq T(k - |S^{1}_B|) + \ldots + T(k - |S^{q}_B|)$,
so we have
\begin{equation}
\label{inequality3}
T(k) \leq |\mathcal{C}_r| \cdot T(k - r) + \ldots + |\mathcal{C}_h| \cdot T(k - h).
\end{equation}

Since Branching Rule 1 goes two branches, we must have
$$\frac{|\mathcal{C}_r|}{2^r} + \ldots + \frac{|\mathcal{C}_h|}{2^h} = 1.$$

By the well-known methods in algorithm analysis~\cite{fptbook}, it can be derived that the
characteristic polynomial for the recurrence relation (\ref{inequality3}) is
$x^h - (|\mathcal{C}_r| \cdot x^{h - r} + \ldots + |\mathcal{C}_{h-1}| \cdot x + |\mathcal{C}_h|)$,
whose unique positive root is $x = 2$ that is also the root of the characterisitc polynomial of the
recurrence relation $T(k) = 2T(k - 1)$. In conclusion, the $m$-BR-process on $(F_1, \ldots, F_m; k)$
satisfies the recurrence relation $T(k) \leq 2 T(k-1)$.
\end{proof}
\end{theorem}

Corollary~\ref{cor:cor1} and Theorem~\ref{theo:m-BR-process} are valid for any instance
$(F_1, \ldots, F_m; k)$ of {\sc hMaf}. In the following, we study the $m$-BR-process on
instances satisfying the {\it 2-edge distance property}, whose definition is given as
follows.

An instance $(F_1, \ldots, F_m; k)$ of the {\sc Maf} problem (hard or soft) satisfies the {\it 2-edge
distance property} if (1) there is an instance $(\mathcal{F}_1, \ldots, \mathcal{F}_m; \mathcal{K})$ that satisfies
the label-set isomorphism property, and an edge-set $E$ that contains at most one edge
in each $\mathcal{F}_i$ such that $(\mathcal{F}_1, \ldots, \mathcal{F}_m; \mathcal{K}-1) \setminus E = (F_1, \ldots, F_m; k)$,
where $\mathcal{K} = k+1$; and (2) there are two edges $e_i = [v_i^1,v_i^2]$ and
$e_j = [v_j^1,v_j^2]$ of $E$ that are in $\mathcal{F}_i$ and $\mathcal{F}_j$ ($i < j$), respectively, such
that $L(v_i^2) \neq L(v_j^2)$ (the vertices $v_i^1$ and $v_j^1$ are the parents of $v_i^2$
and $v_j^2$, respectively).

\begin{lemma}
\label{lem:speical equal extension}
For an instance $I^* = (F^*_1, \ldots, F^*_m; k^*)$ obtained by the $m$-BR-process on
an instance $I = (F_1, \ldots, F_m; k)$ of the {\sc hMaf} problem that satisfies the 2-edge distance property,
$Ord(F^*_1) \geq |S_B| + 1 + Ord_{\max}(F_1, \ldots, F_m; k)$, where $S_B$ is the set of
edges  removed by Branching Rule 1 in the $m$-BR-process from $I$ to $I^*$.

\begin{proof}
Let $(\bar{F}_1, \ldots, \bar{F}_m; \bar{k})$ be the instance obtained by repeatedly
applying Reduction Rule 1 on $(F_1, \ldots, F_m; k)$ until the rule is unapplicable. If
$Ord_{\max}(\bar{F}_1, \ldots, \bar{F}_m; \bar{k}) \geq Ord_{\max}(F_1, \ldots, F_m;k)+1$,
then the lemma holds true because of Corollary~\ref{cor:cor1}. Thus, in the following
discussion, we analyze the case
$Ord_{\max}(\bar{F}_1, \ldots, \bar{F}_m; \bar{k}) = Ord_{\max}(F_1, \ldots, F_m; k)$.

Let $E'$ be a subset of $E$ with the minimum size such that $\{e_i,e_j\} \subseteq E'$, and
$(\bar{F}_1, \ldots, \bar{F}_m; \bar{k})$ can be obtained by applying Reduction
Rule 1 on $(\mathcal{F}_1, \ldots, \mathcal{F}_m; \mathcal{K}-1) \setminus E' = (\widetilde{F}_1, \ldots, \widetilde{F}_m; k)$
until it is unapplicable. It is easy to see that the collection of instances obtained by the
$m$-BR-process on $(F_1, \ldots, F_m; k)$ is the same as that obtained by the $m$-BR-process on
$(\widetilde{F}_1, \ldots, \widetilde{F}_m; k)$. Thus in the following discussion, we analyze the
$m$-BR-process on $(\widetilde{F}_1, \ldots, \widetilde{F}_m; k)$.

Assume the first 2-BR-process in the $m$-BR-process on $(\widetilde{F}_1, \ldots, \widetilde{F}_m; k)$
is on $\widetilde{F}_i = F_i$ and $\widetilde{F}_j = F_j$, which have the largest orders among the
$X$-forests in $(\widetilde{F}_1, \ldots, \widetilde{F}_m; k)$. If we can construct two connected
label-pairs for $e_i$ and $e_j$ respectively, then we can regard them as the two edges removed by
Branching Rule 1 during the process from $(F^1_1, \ldots, F^1_m; k^1)$ to $(F^*_1, \ldots, F^*_m; k^*)$,
where $(F^1_1, \ldots, F^1_m; k^1)$ is the instance obtained by removing edges in $E' \setminus \{e_i,e_j\}$
from $(\mathcal{F}_1, \ldots, \mathcal{F}_m; \mathcal{K})$. Then by combining Lemma~\ref{lem:equal}, Corollary~\ref{cor:cor1},
and the fact that $Ord_{\max}(F^1_1, \ldots, F^1_m; k^1) +1 \geq Ord_{\max}(F_1, \ldots, F_m; k)$,
we can easily get $Ord(F^*_1) \geq |S_B| + 2 + Ord_{\max}(F^1_1, \ldots, F^1_m; k^1) \geq
|S_B| + 1 + Ord_{\max}(F_1, \ldots, F_m; k)$.

Now we explain how to construct the connected label-pairs for $e_i$ and $e_j$. Let
$(F^2_1, \ldots, F^2_m; k^2)$ be the instance obtained by applying Reduction Rule 1 on
$(F^1_1, \ldots, F^1_m; k^1)$ until it is unapplicable. For the edge $e_i$ in
$(F^2_1, \ldots, F^2_m; k^2)$, we can get a connected label-pair by comparing $F^2_i$
and $F^2_j$. Let $(F^3_1, \ldots, F^3_m; k^3)$ be the instance obtained by applying Reduction
Rule 1 on $(F^2_1, \ldots, F^2_i \setminus \{e_i\}, \ldots, F^2_m; k^2)$ until it is unapplicable.
Then for the edge $e_j$ in $(F^3_1, \ldots, F^3_m; k^3)$ (since $L(v_i^2) \neq L(v_j^2)$,
$e_j$ is in $(F^3_1, \ldots, F^3_m; k^3)$), we can also get a connected label-pair by
comparing $F^3_i$ and $F^3_j$.
\end{proof}
\end{lemma}

Let $\mathcal{C} = \{(F^1_1, \ldots, F^1_m; k^1), \ldots, (F^q_1, \ldots, F^q_m; k^q)\}$
be the collection of instances that are obtained by the $m$-BR-process on $(F_1, \ldots, F_m; k)$
satisfying the 2-edge distance property, and let $S^p_B$, $1 \leq p \leq q$, be the set of edges
removed by Branching Rule 1 during the $m$-BR-process from $(F_1, \ldots, F_m; k)$ to
$(F^p_1, \ldots, F^p_m;k^p)$. Let
$r = \min\{|S^{1}_B|, \ldots, |S^{q}_B|\}$ and $h = \max\{|S^{1}_B|, \ldots, |S^{q}_B|\}$,
and let $\mathcal{C}_l$ be the subset of $\mathcal{C}$ that contains all
$(F^p_1, \ldots, F^p_m; k^p)$ in $\mathcal{C}$ such that $|S^p_B| = l$, for any $r \leq l \leq h$.
Combining Theorem~\ref{theo:m-BR-process} and Lemma~\ref{lem:speical equal extension}, we
get the following two theorems.

\begin{theorem}
\label{theo:special m-BR-process 0}
For the $m$-BR-process on $(F_1, \ldots, F_m; k)$ satisfying the 2-edge distance property, if
Branching Rule 1 is not applied, then the parameter of the unique instance obtained by it has
value not greater than $k-1$.
\end{theorem}

\begin{theorem}
\label{theo:special m-BR-process 1}
For the $m$-BR-process on $(F_1, \ldots, F_m; k)$ satisfying the 2-edge distance property, if
Branching Rule 1 is applied, then it satisfies the recurrence relation
$T(k) \leq |\mathcal{C}_r| \cdot T(k - r-1) + \ldots + |\mathcal{C}_h| \cdot T(k - h-1)$, where
$\frac{|\mathcal{C}_r|}{2^r} + \ldots + \frac{|\mathcal{C}_h|}{2^h} = 1$.

\begin{proof}
For any instance $(F^{p}_1, \ldots, F^{p}_m; k^{p})$ in $\mathcal{C}$, by
Lemma~\ref{lem:speical equal extension}, $k^p \leq k - |S^p_B|-1$. Thus, $T(k^p) \leq T(k - |S^p_B|-1)$
for all $1 \leq p \leq q$. This gives
$T(k) = T(k^1) + \ldots + T(k^q) \leq T(k - |S^{1}_B|-1) + \ldots + T(k - |S^{q}_B|-1)$,
so
\begin{equation}
\label{inequality7}
T(k) \leq |\mathcal{C}_r| \cdot T(k - r-1) + \ldots + |\mathcal{C}_h| \cdot T(k - h-1).
\end{equation}

Similar to the analysis for Theorem~\ref{theo:m-BR-process}, we have
$\frac{|\mathcal{C}_r|}{2^r} + \ldots + \frac{|\mathcal{C}_h|}{2^h} = 1$.
\end{proof}
\end{theorem}

\section{Analysis for Maximal Sibling Set}

In this section, we assume that the instance $(F_1, \ldots, F_m; k)$ of the {\sc hMaf}
problem satisfies the label-set isomorphism property. We start with the following reduction rule.

\medskip

\noindent{\bf Reduction Rule 2.} If there is a subset $S$ of $X$ that is an MSS in
$F_i$ for all $i$, then for all $1 \leq i \leq m$, group $T_{F_i}[S]$
into an un-decomposable structure, and mark it with the same new label, here $T_{F_i}[S]$
is the subtree in $F_i$ rooted at the common parent of the labels in $S$.

\medskip

Under the condition of Reduction Rule 2, $T_{F_1}[S]$, $\ldots,$ $T_{F_m}[S]$ are
isomorphic. It is easy to see that all labels in $S$ are in the same connected
component of every MAF $F^*$ for the $X$-forests in $(F_1, \ldots, F_m; k)$, and
$T_{F^*}[S] = T_{F_1}[S] = \cdots = T_{F_m}[S]$. Thus, $T_{F_1}[S]$, $\ldots$,
$T_{F_m}[S]$ remain unchanged in the further processing, so we can treat them as
an un-decomposable structure.

To implement Reduction Rule 2, we simply remove all labels in $S$, label their common
parent with a new label $l_S$, and replace the label-set $X$ with a new label-set
$(X \setminus S) \cup \{l_S\}$. We have the following lemma.

\begin{lemma}
\label{lem:Reduction Rule 2}
For the instance $(F'_1,F'_2,\ldots,F'_m;k)$ that is obtained by applying Reduction Rule 2 on $(F_1,F_2,\ldots,F_m;k)$ with grouping MSS $S$, $(F_1,F_2,\ldots,F_m;k)$ has the same collection of solutions with it.

\begin{proof}
By above analysis, for each MAF $F'$ for the forests in $(F'_1,F'_2,\ldots,F'_m;k)$, a corresponding MAF for the forests in $(F_1,F_2,\ldots,F_m;k)$ can be constructed by replacing the label $\overline{S}$ in $F'$ with the subtree $T_{F_1}[S]$; for each MAF $F$ for the forests in $(F_1,F_2,\ldots,F_m;k)$, a corresponding MAF for the forests in $(F'_1,F'_2,\ldots,F'_m;k)$ can be constructed by replacing the subtree $T_{F_1}[S]$ in $F$ with the label $\overline{S}$.

Because of the bijective relation between the MAFs for the forests in $(F_1,F_2,\ldots,F_m;k)$ and that for the forests in $(F'_1,F'_2,\ldots,F'_m;k)$, we simply say that $(F'_1,F'_2,\ldots,F'_m;k)$ and $(F_1,F_2,\ldots,F_m;k)$ have the same collection of solutions.
\end{proof}
\end{lemma}

\begin{lemma}
\label{lem:no-MSS}
For any instance $(F_1,F_2,\ldots,F_m;k)$ of the {\sc Maf} problem (hard or soft) that satisfies the label-set isomorphism property, if $F_i$ has no MSS, $1 \leq i \leq m$, then $F_i$ is the unique MAF for the $X$-forests in $(F_1,F_2,\ldots,F_m;k)$.

\begin{proof}
If $F_i$ has no MSS, then $F_i$ has at most one edge (the edge between label $\rho$ and some label in $X \setminus \{\rho\}$). Combining the fact that $(F_1,F_2,\ldots,F_m;k)$ satisfies the label-set isomorphism property, we can easily get that all $X$-forests in $(F_1,F_2,\ldots,F_m;k)$ are isomorphic, and $F_i$ is the unique MAF for the $X$-forests in $(F_1,F_2,\ldots,F_m;k)$.
\end{proof}
\end{lemma}

In the following discussion, we assume that Reduction Rule 2 is unapplicable on the
instance $I = (F_1, \ldots, F_m; k)$. An MSS $S$ of an $X$-forest $F_h$ in $I$ is a
{\it minimum MSS} in $I$ if no $X$-forest in $I$ has an MSS whose size is smaller
than that of $S$. By Lemma~\ref{lem:no-MSS} and without loss of generality, we assume
that $F_1$ has a minimum MSS $S$ in $(F_1, \ldots, F_m; k)$. Following the discussion
on Case 1 in the previous section, we consider the remaining cases.

\subsection{Case 2: $|S| = 2$}

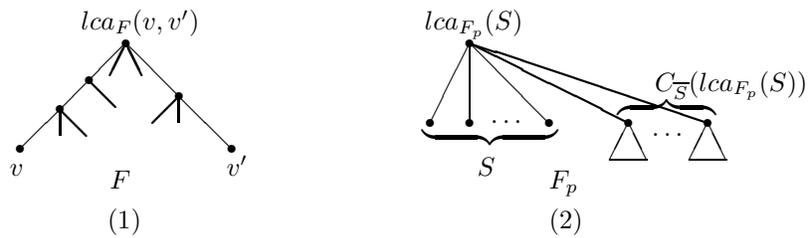
\begin{figure}[h]
\begin{center}
\begin{picture}(300,80)

\put(20,-5){\begin{picture}(0,0)
 \put(30,70){\circle*{3}}
 \put(13,75){\small $lca_{F}(v,v')$}
 \put(30,70){\line(-1,-1){40}}
 \thicklines
 \put(30,70){\line(-1,-2){6}}
 \put(30,70){\line(1,-2){6}}
 \put(16,56){\circle*{3}}
 \put(16,56){\line(1,-1){10}}
 \put(5,45){\circle*{3}}
 \put(5,45){\line(1,-1){10}}
 \put(5,45){\line(0,-1){10}}
  \thinlines

 \put(30,70){\line(1,-1){40}}
 \put(50,50){\circle*{3}}
 \thicklines
 \put(50,50){\line(-1,-1){10}}
 \put(50,50){\line(0,-1){10}}
 \thinlines
 \put(-10,30){\circle*{3}}
 \put(-14,20){\small $v$}
 \put(70,30){\circle*{3}}
 \put(68,20){\small $v'$}
  \put(24,15){\small $F$}
  \put(23,0){\small (1)}
\end{picture}}

\put(150,-5){\begin{picture}(0,0)
 \put(30,70){\circle*{3}}
 \put(13,75){\small $lca_{F_p}(S)$}
 \put(30,40){\circle*{3}}
 \put(15,40){\circle*{3}}
 \put(60,40){\circle*{3}}
 \put(38,40){\small $\ldots$}
 \put(30,70){\line(0,-2){30}}
 \put(30,70){\line(-1,-2){15}}
 \put(30,70){\line(1,-1){30}}

 \put(12,38){$\underbrace{\rule{1.8cm}{0cm}}$}
 \put(33,20){\small $S$}

  \thicklines
 \put(30,70){\line(2,-1){60}}
  \thinlines
 \put(90,40){\circle*{3}}

 \put(90,40){\line(-1,-2){7}}
 \put(90,40){\line(1,-2){7}}
 \put(83,26){\line(1,0){14}}
 \put(99,35){\small $\ldots$}

 \thicklines
 \put(30,70){\line(3,-1){90}}
 \thinlines
 \put(120,40){\circle*{3}}

 \put(120,40){\line(-1,-2){7}}
 \put(120,40){\line(1,-2){7}}
 \put(113,26){\line(1,0){14}}
 \put(86,42){$\overbrace{\rule{1.3cm}{0cm}}$}
 \put(100,52){\small $C_{\overline{S}}(lca_{F_p}(S))$}
 \put(60,15){\small $F_p$}
 \put(60,0){\small (2)}
\end{picture}}
\end{picture}
\end{center}
\vspace*{-5mm}
\caption{(1) The set $E_F(v,v')$, which consists of the bold edges.
(2) The $X$-forest $F_p$ for Case 3.1, where the set $E_{F_p}^{O}(S)$ consists of
the bold edges and the triangles are subtrees.}
\label{fig:|S|>=2}
\vspace*{-5mm}
\end{figure}

Given two vertices $v$ and $v'$ that are in the same connected component of an
$X$-forest $F$, denote by $lca_F(v,v')$ the least common ancestor of $v$ and $v'$
in $F$. Let $P = \{v,v_1,\ldots,v_r,v'\}$ be the path connecting $v$ and $v'$ in $F$.
Denote by $E^1_{F}(v,v')$ the set containing all edges that are not on the path $P$,
but are incident to the vertices in the set $P \setminus \{v,v',lca_F(v,v')\}$. Denote
by $E^2_{F}(v,v')$ the set containing all edges that are incident to $lca_F(v,v')$,
except the edges on $P$ and the edge between $lca_F(v,v')$ and its parent. Let
$E_{F}(v,v') = E^1_{F}(v,v') \cup E^2_{F}(v,v')$. See Figure~\ref{fig:|S|>=2}(1) for
an illustration. In this subsection, we assume that the minimum MSS is $S = \{a,b\}$,
which is in $F_1$.

\medskip

\noindent{\bf Case 2.1.} There is an $X$-forest $F_p$ such that $|E_{F_p}(a,b)| \geq 2$.

\medskip

\noindent{\bf Branching Rule 2.1.} Branch into three ways: [1] remove the edge incident
to $a$ in all $X$-forests; [2] remove the edge incident to $b$ in all $X$-forests;
[3] remove the edges in $E_{F_i}(a,b)$ for all $2 \leq i \leq m$.

\begin{lemma}
\label{lem:Branching Rule 2.1}
Branching Rule 2.1 is safe, and satisfies the recurrence relation
$T(k) \leq 2 T(k-1) + T(k-2)$.

\begin{proof}
Let $F^*$ be an MAF for the $X$-forests in $(F_1, \ldots, F_m; k)$. There are three
possible cases for $a$ and $b$ in $F^*$.

(1) Label $a$ is a single-vertex tree in $F^*$. Then the first branch of Branching
Rule 2.1 is correct. Since $Ord(F_1) = \cdots = Ord(F_m)$ and each $X$-forest in the
new instance obtained by the first branch has order $Ord(F_1) + 1$, the value of the
parameter in the new instance is $k - 1$.

(2) Label $b$ is a single-vertex tree in $F^*$. Similarly to the analysis for
case (1): the second branch of Branching Rule 2.1 is correct, and the value of the
parameter in the new instance is $k-1$.

(3) Labels $a$ and $b$ are in the same connected component of $F^*$. Since $a$ and
$b$ are a sibling-pair in $F_1$, they are also a sibling-pair in $F^*$. In order
to make $a$ and $b$ a sibling-pair in $F_i$, for each $i \geq 2$, the edges in
$E_{F_{i}}(a,b)$ should be removed. Thus, the third branch of Branching Rule 2.1 is
correct. Since $E_{F_{p}}(a,b)$ is an ee-set in $F_p$ and $|E_{F_{p}}(a,b)| \geq 2$,
$Ord(F_p \setminus E_{F_{p}}(a,b)) \geq Ord(F_p) + 2$ and $k' \leq k - 2$, where $k'$
is the parameter of the new instance
$(F_1, F_2 \setminus E_{F_{2}}(a,b), \ldots, F_m \setminus E_{F_{m}}(a,b); k')$.

Combining the above discussion shows that the recurrence relation of Branching Rule 2.1
is $T(k) = 2T(k-1) + T(k') \leq 2T(k-1) + T(k-2)$.
\end{proof}
\end{lemma}

\noindent{\bf Case 2.2.} For all $2 \leq i \leq m$, $|E_{F_i}(a,b)| \leq 1$.

\medskip

Let $\mathcal{C}$ be the collection of the $X$-forests in $(F_1, \ldots, F_m; k)$
such that for each $F$ in $\mathcal{C}$, $|E_{F}(a,b)| = 1$. Note that
$\mathcal{C} \neq \emptyset$ -- otherwise, labels $a$ and $b$ would be a sibling-pair
in all forests so Reduction Rule 2 would be applied.

\medskip

\noindent{\bf Case 2.2.1} $L(lca_{F}(a,b)) = L(lca_{F'}(a,b))$ for all $X$-forests
$F$ and $F'$ in $\mathcal{C}$.

\begin{lemma}
\label{lem:sibling-2}
For Case 2.2.1, there exists an MAF for the $X$-forests in $(F_1, \ldots, F_m; k)$, in
which labels $a$ and $b$ are a sibling-pair.

\begin{proof}
Let $F^*$ be an MAF for the $X$-forests in $(F_1, \ldots, F_m; k)$. Pick an $X$-forest
$F_p$ in $\mathcal{C}$ and assume $E_{F_p}(a,b) = \{e\}$. Let $e_a$ and $e_b$ be the edges
incident to $a$ and $b$ in $F_p$, respectively.

If $a$ and $b$ are in the same connected component of $F^*$, then $a$ and $b$ are a
sibling-pair in $F^*$ since $a$ and $b$ are a sibling-pair in $F_1$, so the lemma is
proved. Thus, we assume that $a$ and $b$ are not in the same connected component of
$F^*$, i.e., at least one of $a$ and $b$ is a single-vertex tree in $F^*$. Without
loss of generality, assume that $a$ is a single-vertex tree in $F^*$. There are two
cases for $b$ in $F^*$.

(1) Label $b$ is a single-vertex tree in $F^*$. There must be a connected component
of $F^*$ that contains labels in $L(lca_{F_p}(a,b)) \setminus \{a,b\}$ as well as
labels in $X \setminus L(lca_{F_p}(a,b))$ -- otherwise, by attaching $a$ to the
single-vertex tree $b$ in $F^*$ to make $a$ and $b$ a sibling-pair, an agreement forest
for the $X$-forests in $(F_1, \ldots, F_m; k)$ with an order smaller than that of $F^*$
would be made, contradicting the fact that $F^*$ is an MAF for $(F_1, \ldots, F_m; k)$.
Thus, the edge $e$ and the edge $e_{lca}$ between $lca_{F_p}(a,b)$ and its parent are
in $F^*$.

Let $E$ be an ee-set of $F_p$ such that $F_p \setminus E = F^*$ and $e_a, e_b \in E$. By
the above analysis, the edges $e$ and $e_{lca}$ cannot be in $E$. We can get an ee-set
$E'$ of $F_p$ by replacing $e_a$ and $e_b$ with $e$ and $e_{lca}$ in $E$, and easily verify
(recall $L(lca_{F}(a,b)) = L(lca_{F'}(a,b))$ for all $F$ and $F'$ in $\mathcal{C}$)
that $F_p \setminus E'$ is also an MAF for $(F_1, \ldots, F_m; k)$, in which $a$ and
$b$ are a sibling-pair.

(2) Label $b$ is not a single-vertex tree in $F^*$. If the edge $e$ is not in $F^*$,
then by attaching $a$ to the middle of the edge incident to $b$ in $F^*$ to make $a$ and
$b$ a sibling-pair, we would get an agreement forest for $(F_1, \ldots, F_m; k)$ whose
order is smaller than that of the MAF $F^*$, deriving a contradiction. Thus, the edge
$e$ must be in $F^*$.

Let $E$ be an ee-set of $F_p$ such that $F_p \setminus E = F^*$ and $e_a \in E$. By the
above analysis, edge $e$ in not in $E$. We can get an ee-set $E'$ of $F_p$ by replacing
$e_a$ with $e$ in $E$, and again easily verify that $F_p \setminus E'$ is also an MAF for
$(F_1, \ldots, F_m; k)$, in which $a$ and $b$ are a sibling-pair.
\end{proof}
\end{lemma}

\noindent{\bf Reduction Rule 2.2.1.} Under the condition of Case 2.2.1, remove the edge
in $E_{F_i}(a,b)$ for all $2 \leq i \leq m$.

\medskip

Lemma~\ref{lem:sibling-2} immediately implies the following lemma.

\begin{lemma}
\label{lem:Branching Rule 2.2.1}
Reduction Rule 2.2.1 on an instance $I$ of the {\sc hMaf} problem produces an instance
that is a yes-instance if and only if $I$ is a yes-instance.
\end{lemma}

\noindent{\bf Case 2.2.2.} There are $F_p$ and $F_q$ in $\mathcal{C}$ with
$L(lca_{F_p}(a,b)) \neq L(lca_{F_q}(a,b))$.

\medskip

\noindent{\bf Branching Rule 2.2.2.} Under the condition of Case 2.2.2, branch into
three ways: [1] remove the edge incident to $a$ in all $X$-forests; [2] remove the
edge incident to $b$ in all $X$-forests; [3] remove the edge in $E_{F_i}(a,b)$ for
all $2 \leq i \leq m$, and apply $m$-BR-process.

\begin{lemma}
\label{Branching Rule 2.2.2}
Branching Rule 2.2.2 is safe, and satisfies the recurrence relation
$T(k) \leq 2T(k-1) + T(k-2)$.

\begin{proof}
Let $F^*$ be an MAF for the $X$-forests in $(F_1, \ldots, F_m; k)$. There are three
possible cases for $a$ and $b$ in $F^*$.

(1-2) Label $a$ or $b$ is a single-vertex tree in $F^*$. Using the analysis for the
first two cases in the proof of Lemma~\ref{lem:Branching Rule 2.1}, we can derive that
the first two branches of Branching Rule 2.2.2 are correct, and the two branches
construct two new instances whose parameter values are $k-1$.

(3) Labels $a$ and $b$ are in the same connected component of $F^*$. Using the analysis
for the third case in the proof of Lemma~\ref{lem:Branching Rule 2.1}, we can derive that
the third branch of Branching Rule 2.2.2 is correct.

For the new instance $I = (F_1, F_2 \setminus E_{F_{2}}(a,b), \ldots, F_m \setminus E_{F_{m}}(a,b); k')$
obtained by the third branch, we can see that $I$ satisfies the 2-edge distance property and
that $k' = k-1$. The discussion about the $m$-BR-process on $I$ is divided into two subcases.

(3.1) Branching Rule 1 is not applied during the $m$-BR-process on $I$. By
Theorem~\ref{theo:special m-BR-process 0}, only one instance is obtained by the
$m$-BR-process, whose parameter $k''$ has value not larger than $k'-1 = k-2$. Thus,
in this subcase, the recurrence relation of Branching Rule 2.2.2 is
$T(k) = 2T(k-1) + T(k'') \leq 2T(k-1) + T(k-2)$.

(3.2) Branching Rule 1 is applied during the $m$-BR-process on $I$. By
Theorem~\ref{theo:special m-BR-process 1},
$T(k') \leq |\mathcal{C}_r| \cdot T(k' - r-1) + \ldots + |\mathcal{C}_h| \cdot T(k' - h-1)$,
where $k' = k-1$ and $\frac{|\mathcal{C}_r|}{2^r} + \ldots + \frac{|\mathcal{C}_h|}{2^h} = 1$.
Thus, in this subcase, the recurrence relation of Branching Rule 2.2.2 is
\[ T(k) \leq 2T(k-1) + |\mathcal{C}_r| \cdot T(k - r-2) + \ldots + |\mathcal{C}_h| \cdot T(k - h-2).\]
The characteristic polynomial of the above recurrence relation is
$p(x) = x^{h+2} - 2 x^{h+1} - |\mathcal{C}_r| \cdot x^{h-r} - \ldots - |\mathcal{C}_h|$.
Since $p(2) < 0$ and $p(1+\sqrt{2}) > 0$, the unique positive root of $p(x)$ has its value
bounded by $1+\sqrt{2}$. Therefore, if Branching Rule 1 is applied during the $m$-BR-process
on $I$, then Branching Rule 2.2.2 satisfies the recurrence relation $T(k) \leq 2T(k-1) + T(k-2)$,
whose characteristic polynomial has its unique positive root $1+\sqrt{2}$.

Summarizing these discussions, we conclude that the recurrence relation of Branching
Rule 2.2.2 satisfies $T(k) \leq 2T(k-1) + T(k-2)$.
\end{proof}
\end{lemma}

\subsection{Case 3: $|S| \geq 3$}

For an $X$-forest $F$ in which the labels in $S$ are in the same connected component,
denote by $lca_{F}(S)$ the least common ancestor of the labels in $S$ in $F$, and
denote by $C_{\overline{S}}(lca_{F}(S))$ the set containing all children of
$lca_{F}(S)$ in $F$ that are not labels in $S$.

\medskip

\noindent{\bf Case 3.1.} All labels in $S$ are siblings in $F_i$, for all
$2 \leq i \leq m$.

\medskip

For Case 3.1, there exists at least one $X$-forest $F_p$ in $(F_1, \ldots, F_m; k)$
such that $C_{\overline{S}}(lca_{F_p}(S)) \neq \emptyset$ -- otherwise, $S$ is an MSS
in all $X$-forests in the instance so that it could be grouped by Reduction Rule 2.
Denote by $E^O_{F_i}(S)$ the set containing all edges between $lca_{F_i}(S)$ and the
vertices in $C_{\overline{S}}(lca_{F_i}(S))$, for  $2 \leq i \leq m$. See
Figure~\ref{fig:|S|>=2}(2) for an illustration. Note that for each vertex
$v \in C_{\overline{S}}(lca_{F_i}(S))$, we have $L(v) \cap S = \emptyset$.

\medskip

\noindent {\bf Branching Rule 3.1.} Branch into two ways: [1] remove the edges incident
to the labels of $S \setminus \{x\}$ in all $X$-forests, where $x$ is an arbitrary label
of $S$; [2] remove the edges in $E^{O}_{F_{i}}(S)$ for all $2 \leq i \leq m$.

\begin{lemma}
\label{Branching Rule 3.1}
Branching Rule 3.1 is safe, and satisfies the recurrence relation $T(k) \leq T(k-1) + T(k-2)$.

\begin{proof}
Let $F^*$ be an MAF for the $X$-forests in $(F_1, \ldots, F_m; k)$. There are two possible
cases for $S$ in $F^*$.

(1) There is a label in $S$ that is a single-vertex tree in $F^*$. In this case, we first
show that there is at most one label in $S$ that is not a single-vertex tree in $F^*$.
Suppose that labels $x_1$ and $x_2$ in $S$ are not single-vertex trees in $F^*$, and label
$x_3$ in $S$ is a single-vertex tree in $F^*$. Since $x_1$ and $x_2$ are siblings in $F_1$,
they are also siblings in $F^*$. By attaching $x_3$ to the common parent of $x_1$ and $x_2$
in $F^*$, $x_1$, $x_2$, and $x_3$ become siblings, which would result in an agreement forest
for $(F_1, \ldots, F_m; k)$ whose order is smaller than that of the MAF $F^*$. This
contradiction shows that at most one label in $S$ is not a single-vertex tree in $F^*$.

Suppose that the label $x$ in $S$ is not a single-vertex tree in $F^*$. By symmetry of the labels
in $S$, there is another MAF $F'$ for $(F_1, \ldots, F_m; k)$ such that $x$ is a single-vertex
tree in $F'$ and some label $x'$ of $S \setminus \{x\}$ is not a single-vertex tree in $F'$.
Thus, the first branch of Branching Rule 3.1 is correct that arbitrarily picks a label $x$ in $S$,
and removes all edges incident to the labels in $S \setminus \{x\}$ in all $X$-forests. Since
$Ord(F_1) = \ldots = Ord(F_m)$  and each $X$-forest in the new instance has order
$Ord(F_1) + |S|-1$, the value of the parameter in the new instance is $k - |S|+1 \leq k-2$.

(2) No label in $S$ is a single-vertex tree in $F^*$. Since $S$ is an MSS in $F_1$, $S$ is
an MSS in $F^*$. Thus, the second branch of Branching Rule 3.1 that removes all edges
in $E^O_{F_i}(S)$ for all $2 \leq i \leq m$ is correct. Suppose that
$|E^O_{F_p}(S)| = \max \{|E^O_{F_2}(S)|, \ldots, |E^O_{F_m}(S)|\}$. Then
$|E^O_{F_p}(S)| \geq 1$ and $E^O_{F_p}(S)$ is an ee-set of $F_p$. Thus
$F_p \setminus E^O_{F_p}(S)$ has the maximum order among all $X$-forests in
$(F_1, F_2 \setminus E^O_{F_2}(S), \ldots, F_m \setminus E^O_{F_m}(S); k')$,
where $k' = k - |E^O_{F_p}(S)|$.

In conclusion, the recurrence relation of Branching Rule 3.1 is
$T(k) = T(k- |S| + 1) + T(k - |E^O_{F_p}(S)|)  \leq T(k - 2) + T(k - 1)$.
\end{proof}
\end{lemma}

\noindent{\bf Case 3.2.} There exists an $X$-forest $F_p$ and two labels $x_1$ and $x_2$ in
$S$ such that $|E^1_{F_p}(x_1, x_2)| \geq 2$.

\medskip

\noindent {\bf Branching Rule 3.2.} Branch into three ways: [1] remove the edge incident to
$x_1$ in all $X$-forests; [2] remove the edge incident to $x_2$ in all $X$-forests; [3] remove
the edges in $E^1_{F_i}(x_1,x_2)$ for all $2 \leq i \leq m$.

\begin{lemma}
\label{Branching Rule 3.2}
Branching Rule 3.2 is safe, and satisfies the recurrence relation $T(k) \leq 2T(k-1) + T(k-2)$.

\begin{proof}
Let $F^*$ be an MAF for the $X$-forests in $(F_1, \ldots, F_m; k)$. There are three possible
cases for the labels $x_1$ and $x_2$ in $F^*$.

(1-2) Label $x_1$ or $x_2$ is a single-vertex tree in $F^*$. Using the analysis for the first
two cases in the proof of Lemma~\ref{lem:Branching Rule 2.1}, we can get that the first
two branches of Branching Rule 3.2 are correct, and the two branches construct two new
instances whose parameter values are $k-1$.

(3) Labels $x_1$ and $x_2$ are in the same connected component in $F^*$. Since $x_1$
and $x_2$ are siblings in $F_1$, $x_1$ and $x_2$ are also siblings in $F^*$. In order to make
$x_1$ and $x_2$ siblings in $F_i$, for each $i \geq 2$, the edges in $E^1_{F_i}(x_1,x_2)$
should be removed (note that the edges in $E^2_{F_i}(x_1, x_2)$ cannot be removed in
this case). Thus, the third branch of Branching Rule 3.2 is correct.

Let $p$ satisfy $|E^1_{F_p}(x_1,x_2)|=\max \{|E^1_{F_2}(x_1,x_2)|,\ldots,|E^1_{F_m}(x_1,x_2)|\}$.
Since $|E^1_{F_p}(x_1,x_2)| \geq 2$ and $E^1_{F_p}(x_1,x_2)$ is an ee-set of $F_p$,
$Ord(F_p \setminus E^1_{F_p}(x_1, x_2)) \geq Ord(F_p) + 2$, and $k' \leq k - 2$, where $k'$ is
the parameter of the new instance
$(F_1, F_2 \setminus E^1_{F_2}(x_1, x_2), \ldots, F_m \setminus E^1_{F_m}(x_1, x_2); k')$.

In conclusion, the recurrence relation of Branching Rule 3.2 is
$T(k) = 2T(k-1) + T(k') \leq 2T(k-1) + T(k-2)$.
\end{proof}
\end{lemma}

\noindent{\bf Case 3.3.} For any $X$-forest $F$ in the instance and any two labels $x$ and
$x'$ in $S$, $|E^1_{F}(x,x')| \leq 1$.
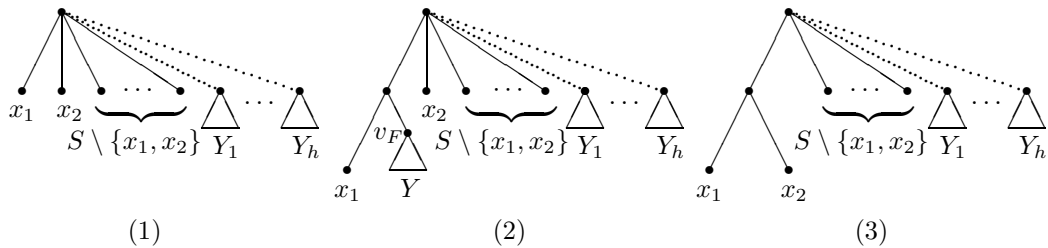
\begin{figure}[h]
\begin{center}
\begin{picture}(330,85)
\put(-40,0){\begin{picture}(0,0)
 \put(30,70){\circle*{3}}
 \put(30,40){\circle*{3}}
 \put(15,40){\circle*{3}}
 \put(45,40){\circle*{3}}
 \put(75,40){\circle*{3}}
 \put(30,70){\line(0,-1){30}}
 \put(30,70){\line(-1,-2){15}}
 \put(30,70){\line(1,-2){15}}
 \put(30,70){\line(3,-2){45}}
 \put(10,30){\small $x_1$}
 \put(28,30){\small $x_2$}
 \put(53,40){\small $\ldots$}
 \put(43,35){$\underbrace{\rule{1.2cm}{0cm}}$}
 \put(32,18){\small $S \setminus \{x_1,x_2\}$}
 \put(55,-16){\small (1)}
 \multiput(30,70)(2,-1){30}{\circle*{1}}
 \put(90,40){\circle*{3}}
 \put(90,40){\line(-1,-2){7}}
 \put(90,40){\line(1,-2){7}}
 \put(83,26){\line(1,0){14}}
 \put(87,16){\small $Y_1$}
 \put(99,35){\small $\ldots$}
 \multiput(30,70)(3,-1){30}{\circle*{1}}
 \put(120,40){\circle*{3}}
 \put(120,40){\line(-1,-2){7}}
 \put(120,40){\line(1,-2){7}}
 \put(113,26){\line(1,0){14}}
 \put(117,16){\small $Y_h$}
\end{picture}}

\put(98,0){\begin{picture}(0,0)
 \put(30,70){\circle*{3}}
 \put(30,40){\circle*{3}}
 \put(15,40){\circle*{3}}
 \put(45,40){\circle*{3}}
 \put(75,40){\circle*{3}}
 \put(30,70){\line(0,-1){30}}
 \put(30,70){\line(-1,-2){15}}
 \put(30,70){\line(1,-2){15}}
 \put(30,70){\line(3,-2){45}}
 \put(15,40){\line(-1,-2){15}}
 \put(90,40){\circle*{3}}
 \put(15,40){\line(1,-2){8}}
 \put(10,22){\small $v_F$}
 \put(23,24){\circle*{3}}
 \put(23,24){\line(-1,-2){7}}
 \put(23,24){\line(1,-2){7}}
 \put(16,10){\line(1,0){14}}
 \put(0,10){\circle*{3}}
 \put(-5,0){\small $x_1$}
 \put(20,0){\small $Y$}
 \put(28,30){\small $x_2$}
 \put(55,38){\small $\cdots$}
 \put(45,35){$\underbrace{\rule{1.2cm}{0cm}}$}
 \put(33,18){\small $S \setminus \{x_1,x_2\}$}
 \put(55,-16){\small (2)}
 \multiput(30,70)(2,-1){30}{\circle*{1}}
 \put(90,40){\circle*{3}}
 \put(90,40){\line(-1,-2){7}}
 \put(90,40){\line(1,-2){7}}
 \put(83,26){\line(1,0){14}}
 \put(87,16){\small $Y_1$}
 \put(99,35){\small $\ldots$}
 \multiput(30,70)(3,-1){30}{\circle*{1}}
 \put(120,40){\circle*{3}}
 \put(120,40){\line(-1,-2){7}}
 \put(120,40){\line(1,-2){7}}
 \put(113,26){\line(1,0){14}}
 \put(117,16){\small $Y_h$}
\end{picture}}

\put(235,0){\begin{picture}(0,0)
 \put(30,70){\circle*{3}}
 \put(15,40){\circle*{3}}
 \put(45,40){\circle*{3}}
 \put(75,40){\circle*{3}}
 \put(30,70){\line(-1,-2){15}}
 \put(30,70){\line(1,-2){15}}
 \put(30,70){\line(3,-2){45}}
 \put(15,40){\line(-1,-2){15}}
 \put(15,40){\line(1,-2){15}}
 \put(0,10){\circle*{3}}
 \put(30,10){\circle*{3}}
 \put(-5,0){\small $x_1$}
 \put(27,0){\small $x_2$}
 \put(53,40){\small $\ldots$}
 \put(43,35){$\underbrace{\rule{1.2cm}{0cm}}$}
 \put(32,18){\small $S \setminus \{x_1,x_2\}$}
 \put(55,-16){\small (3)}
 \multiput(30,70)(2,-1){30}{\circle*{1}}
 \put(90,40){\circle*{3}}
 \put(90,40){\line(-1,-2){7}}
 \put(90,40){\line(1,-2){7}}
 \put(83,26){\line(1,0){14}}
 \put(87,16){\small $Y_1$}
 \put(99,35){\small $\ldots$}
 \multiput(30,70)(3,-1){30}{\circle*{1}}
 \put(120,40){\circle*{3}}
 \put(120,40){\line(-1,-2){7}}
 \put(120,40){\line(1,-2){7}}
 \put(113,26){\line(1,0){14}}
 \put(117,16){\small $Y_h$}
\end{picture}}
\end{picture}
\end{center}
\caption{Three possible structures of $T_{F}[L(lca_{F}(S))]$. Labels $x_1$ and $x_2$
are in $S$. The triangles $Y_l$ are subtrees, $1 \leq l \leq h$ (variable $h$ can be
arbitrarily large).}
\label{not greater than 1}
\end{figure}

\begin{lemma}
\label{structure}
Given a subset $S$ of $X$ with $|S| \geq 3$ and an $X$-forest $F$ in which the labels in $S$
are in the same connected component. If $S$ is not an MSS in $F$, and for any two labels
$x$ and $x'$ in $S$, $|E^1_{F}(x,x')| \leq 1$, then $T_{F}[L(lca_{F}(S))]$ is isomorphic to
one of the cases in Figure~\ref{not greater than 1}.

\begin{proof}
For any two vertices $v_1$ and $v_2$ that are in the same connected component of $F$,
let $N_F(v_1,v_2)$ be the number of internal vertices in the path connecting $v_1$ and
$v_2$. For any label $x$ in $S$, $N_F(x,lca_{F}(S)) \leq 1$.

Suppose there are two labels $x_1$ and $x_2$ in $S$ such that
$N_F(x_1,lca_{F}(S)) = N_F(x_2,lca_{F}(S)) = 1$. If $x_1$ and $x_2$ do not have a
common parent in $F$, then $|E^1_{F}(x_1,x_2)|$ would be $2$. Thus, $x_1$ and
$x_2$ have a common parent $p$ in $F$. By the above analysis, we also derive that
for any label $x$ of $S \setminus \{x_1,x_2\}$, if $N_F(x,lca_{F}(S)) = 1$, then $x$
is a sibling with $x_1$ and $x_2$.

Note that there must be a label $x_3$ in $S \setminus \{x_1,x_2\}$ that is not a sibling
with $x_1$ and $x_2$ -- otherwise, $lca_{F}(S)$ and $p$ would be the same vertex,
contradicting the fact that $N_F(x_1,lca_{F}(S)) = 1$. Then it is easy to see that
$N_F(x_3,lca_{F}(S)) = 0$, i.e., $x_3$ is a child of $lca_{F}(S)$. If $p$ has degree larger
than 3, then $|E^1_{F}(x_1,x_3)|$ would be at least $2$. Thus, the common parent $p$
of $x_1$ and $x_2$ has degree exactly 3, so all labels in $S \setminus \{x_1,x_2\}$ are
children of $lca_{F}(S)$. Thus, $T_{F}[L(lca_{F}(S))]$ is isomorphic to
Figure~\ref{not greater than 1}(3).

In case there is only one label $x_1$ in $S$ satisfying $N_F(x_1,lca_{F}(S)) = 1$, similar to
the analysis above, we can show that the parent $p$ of $x_1$ has degree 3, and all labels
in $S \setminus \{x_1\}$ are children of $lca_{F}(S)$. Thus, $T_{F}[L(lca_{F}(S))]$ is
isomorphic to Figure~\ref{not greater than 1}(2).

If no label $x$ in $S$ satisfies $N_F(x,lca_{F}(S)) = 1$, then all labels in $S$ are children of
$lca_{F}(S)$, and $T_{F}[L(lca_{F}(S))]$ is isomorphic to Figure~\ref{not greater than 1}(1).
\end{proof}
\end{lemma}

Since $S$ is a minimum MSS in $(F_1, \ldots, F_m; k)$, in Case 3.3, no $F$ in
$(F_1, \ldots, F_m; k)$ can make $T_{F}[L(lca_{F}(S))]$ isomorphic to
Figure~\ref{not greater than 1}(3). Thus, for Case 3.3, we only need to consider the
structures (1) and (2) in Figure~\ref{not greater than 1}.

Let $\mathcal{C}_1$ be the collection of the $X$-forests in $(F_1, \ldots, F_m;k)$ such
that for each $F$ in $\mathcal{C}_1$, $T_{F}[L(lca_{F}(S))]$ is isomorphic to
Figure~\ref{not greater than 1}(1), and let
$\mathcal{C}_2 = \{F_1, \ldots, F_m\} \setminus \mathcal{C}_1$. If
$\mathcal{C}_1 = \{F_1, \ldots, F_m\}$, then the labels in $S$ are siblings in all $X$-forests
so that this case can be solved by Branching Rule 3.1. Thus, in the following discussion, we
assume $\mathcal{C}_1 \neq \{F_1,F_2,\ldots,F_m\}$, i.e., $\mathcal{C}_2 \neq \emptyset$.

For each $X$-forest $F$ in $\mathcal{C}_2$, let $x_1$ in $S$ satisfy $N_F(x_1,lca_{F}(S)) = 1$.
Denote by $v_F$ the vertex that has a common parent with $x_1$ in $F$, and by $e_F$ the edge
between $v_F$ and its parent in $F$. See Figure~\ref{not greater than 1}(2) for an illustration.
Assume that $S = \{x_1,x_2,\ldots,x_{|S|}\}$.

\medskip

\noindent{\bf Branching Rule 3.3.} Branch into $1+|S|$ ways: [1] remove the edge $e_{F}$ for
each $X$-forest $F$ in $\mathcal{C}_2$; [1+$i$] let $S' = S \setminus \{x_i\}$, $1 \leq i \leq |S|$,
and remove the edges incident to the labels of $S'$ in all $X$-forests.

\begin{lemma}
\label{lem:Branching Rule 3.3}
Branching Rule 3.3 is safe, and satisfies the recurrence relation $T(k) \leq 2T(k-1) + T(k-2)$.

\begin{proof}
Let $F^*$ be an MAF for the $X$-forests in $(F_1, \ldots, F_m; k)$. There are three possible
cases for the labels of $S$ in $F^*$. Let $S'$ be the subset of $S$ in which each label is not a
single-vertex tree in $F^*$.

(1) $|S'| \geq 2$. Pick an $X$-forest $F_p$ in $\mathcal{C}_2$, and assume that $x_1 \in S$
is the grandchild of $lca_{F_p}(S)$. We show that the edge $e_{F_p}$ is not in $F^*$. If
$x_1 \in S'$, then obviously, the first branch of Branching Rule 3.3 is correct. If $x_1 \notin S'$,
then $x_1$ is a single-vertex tree in $F^*$. Suppose that the edge $e_{F_p}$ of $F_p$ is in
$F^*$. By the structure of $F_p$, we derive that there would be at least one label
$l \in L(v_{F_p})$ that is in the same connected component with labels of $S'$ in $F^*$.
Moreover, the label $l$ is a descendant of $lca_{F^*}(S')$, i.e., $l \in L(lca_{F^*}(S'))$ (note
that $l \notin S$). However, since $F^*$ is a subforest of $F_1$, we must have
$L(lca_{F^*}(S')) \subseteq L(lca_{F_1}(S')) = S$. Thus, that edge $e_{F_p}$ of $F_p$ is in
$F^*$ is impossible, and the first branch of Branching Rule 3.3 is correct.

For the instance $(F'_1, \ldots, F'_m; k')$ that is obtained by removing the edge $e_{F}$ for
each $F$ in $\mathcal{C}_2$, we have $k' = k-1$. Thus, for the first branch of Branching
Rule 3.3, we have $T(k') \leq T(k-1)$.

(2) $|S'| = 1$. We branch by removing edges incident to the labels of $S' = S \setminus \{x_i\}$
in all $X$-forests, for all $1 \leq i \leq |S|$. Since $Ord(F_1) = \ldots = Ord(F_m)$,  each
$X$-forest in the new instance obtained by the $(1+i)$-th branch of Branching Rule 3.3
has order $Ord(F_1) + |S| - 1$, for all $1 \leq i \leq |S|$, and the value of the parameter in
the new instance is $k + 1 - |S|$.

(3) $|S'| = 0$, i.e., all labels are single-vertex trees in $F^*$. Apparently, each of the $(1+i)$-th
branch is correct, for $1 \leq i \leq |S|$.

Therefore, the recurrence relation of Branching Rule 3.3 is
$T(k) = T(k') + |S| \cdot T(k+1-|S|) \leq T(k-1) + |S| \cdot T(k+1-|S|)$, where $|S| \geq 3$.

It is easy to verify that the unique positive root of the characteristic polynomial
$x^{|S|-1} - x^{|S|-2} - |S|$ of the above recurrence relation has its value between 1 and
$1+\sqrt{2}$, for any $|S| \geq 3$. Thus, Branching Rule 3.3 satisfies the recurrence relation
$T(k) \leq 2T(k-1) + T(k-2)$, whose characteristic polynomial has its positive root $1+\sqrt{2}$.
\end{proof}
\end{lemma}

\section{Parameterized Algorithm for the {\sc hMaf} Problem}

\begin{figure}[ht]
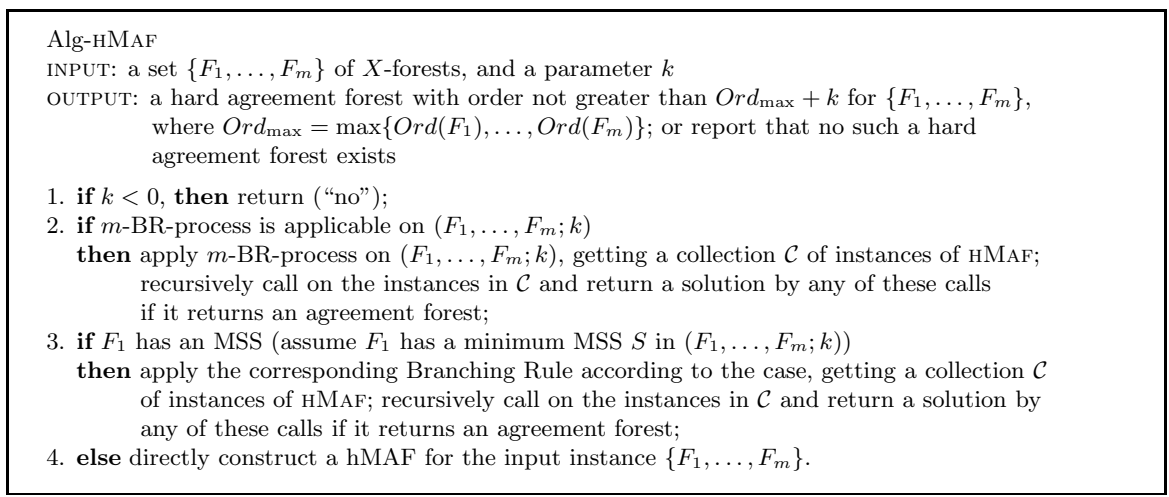

\vspace*{-3mm}
\setbox4=\vbox{\hsize18pc \noindent\strut
\begin{quote}
  \vspace*{-6mm} \footnotesize
\hspace*{-7mm}
Alg-{\sc hMaf}\\
\hspace*{-7mm}  {\sc input}: a set $\{F_1, \ldots, F_m\}$ of $X$-forests, and a parameter $k$\\
  \hspace*{-7mm}
  {\sc output}: a hard agreement forest with order not greater than $Ord_{\max} + k$ for $\{F_1, \ldots, F_m\}$,\\
\hspace*{7mm} where $Ord_{\max} = \max \{Ord(F_1), \ldots, Ord(F_m)\}$; or report that no such a hard \\
\hspace*{7mm} agreement forest exists

  \hspace*{-7mm}
  1.  {\bf if} $k < 0$, {\bf then} return (``no'');\\
  \hspace*{-7mm}
  2.  {\bf if} $m$-BR-process is applicable on $(F_1, \ldots, F_m; k)$\\
  \hspace*{-3mm} {\bf then} apply $m$-BR-process on $(F_1, \ldots, F_m; k)$, getting a collection
              $\mathcal{C}$ of   instances of {\sc hMaf};\\
\hspace*{5.5mm}  recursively call on the instances in $\mathcal{C}$ and return a solution by any of these calls\\
\hspace*{5.5mm}   if it returns an agreement forest; \\
  \hspace*{-7mm}
  3.  {\bf if} $F_1$ has an MSS (assume $F_1$ has a minimum MSS $S$ in $(F_1, \ldots, F_m; k)$)\\
 \hspace*{-3mm} {\bf then} apply the corresponding Branching Rule according to the case, getting a collection $\mathcal{C}$\\
 \hspace*{5.5mm} of instances of {\sc hMaf}; recursively call on the instances in $\mathcal{C}$ and return a solution by\\
 \hspace*{5.5mm} any of these calls if it returns an agreement forest;\\
  \hspace*{-7mm}
  4. {\bf else} directly construct a hMAF for the input instance $\{F_1, \ldots, F_m\}$.
\end{quote} \vspace*{-6mm} \strut} $$\boxit{\box4}$$
\vspace*{-10mm}
\caption{A parameterized algorithm for the {\sc hMaf} problem}
\label{fig:alg-hMaf}
\vspace*{-2mm}
\end{figure}

Our parameterized algorithm for the {\sc hMaf} problem is given in Figure~\ref{fig:alg-hMaf}.

\begin{theorem}
\label{theo:hMaf}
Algorithm  Alg-{\sc hMaf} correctly solves the {\sc hMaf} problem in time $O(2.42^k m^3 n^4)$, where $n$ is the size of the label-set $X$ and $m$ is the number of $X$-forests in the input instance.

\begin{proof}
We first consider the correctness of the algorithm Alg-{\sc hMaf}. Since an agreement forest
$F$ for $\{F_1, \ldots, F_m\}$ is a subforest of each $F_i$, $Ord(F) \geq Ord(F_i)$. If
$k < 0$, then the instance asks for an agreement forest whose order is less than $Ord(F_i)$
for some $F_i$. Apparently, the instance must be a no-instance. Thus, Step 1 of Alg-{\sc hMaf}
is correct. By the discussions given in the previous sections, Steps 2-3 of  Alg-{\sc hMaf} are
also correct. For Step 4 of Alg-{\sc hMaf}, when $F_1$ has no MSS, by Lemma~\ref{lem:no-MSS},
$F_1$ is the unique MAF for $\{F_1, \ldots, F_m\}$. Note that the group operation in Reduction
Rule 2 may change the label-set, but it is straightforward to restore, in linear-time, the
label-set from $F_1$ and get a solution for the original input instance. Therefore, algorithm
Alg-{\sc hMaf} correctly solves the {\sc hMaf} problem.

Now consider the complexity of the algorithm Alg-{\sc hMaf}. For an instance $(F_1, \ldots, F_m; k)$
of the {\sc hMaf} problem, the execution of the algorithm can be depicted as a search tree
$\mathcal{T}$. Each leaf of $\mathcal{T}$ corresponds to a conclusion (either an agreement
forest of order bounded by $Ord_{\max}+k$ or ``no'') of the algorithm. Each internal node in
$\mathcal{T}$ corresponds to a branch for a branching rule used in Steps 2-3 of the algorithm.
We call a path from the root to a leaf in $\mathcal{T}$ a {\it computational path} in the process,
which corresponds to a particular sequence of executions in the algorithm that leads to a conclusion.
The algorithm returns an agreement forest for the original input if and only if there is a computational
path that outputs the forest.

Let $T(k)$ be the number of leaves in $\mathcal{T}$ when the instance parameter is $k$. Then
$T(k)$ satisfies the recurrence relations given for the branching rules discussed in the previous
sections. As discussed, the worst recurrence relation among these recurrence relations is
$T(k) \leq 2T(k-1)+T(k-2)$. By using the standard technique in parameterized
computation~\cite{fptbook}, we get $T(k) \leq O(2.42^k)$, i.e., the search tree $\mathcal{T}$
has $O(2.42^k)$ leaves.

Now we analyze the time spent by a computational path $\mathcal{P}$ between any two consecutive
branches. We consider all possible operations that can be applied on an instance $(F_1',...,F_m';k')$.
If Reduction Rule 1 is applicable, then Reduction Rule 1 should be repeatedly applied until it becomes
unapplicable. Whether Reduction Rule 2 should be applied depends on the test whether the instance
satisfies the label-set isomorphism property.

Without loss of generality, assume that the label-set $X$ is $\{1, 2, \ldots, n\}$. We first
apply a DFS on each $X$-forest $F_i'$ to record $L(v)$ for each vertex $v$ in $F_i'$. This
takes time $O(mn)$, where we can also get the order of $F_i'$, and the label-set of each
connected component of $F_i'$. Then for each connected component $C$ of $F_i'$, we sort
the labels in $L(C)$. This takes time $O(mn \log n)$.

The 3-stage way to decide whether Reduction Rule 1 is applicable on a vertex $v$ in
$F_i'$ relative to $F_j'$, $i < j$, is given as follows. Stage-1: construct a collection
$\mathcal{S}$ of connected components of $F_j'$ such that each $C'$ in $\mathcal{S}$
satisfies $L(C') \cap L(v) \neq \emptyset$. This takes time $O(n^2)$. Stage-2: check if
$L(v) = L(\mathcal{S})$, where $L(\mathcal{S})$ is the union of the label-sets of the
connected components in $\mathcal{S}$. We can first sort the labels in $L(\mathcal{S})$
and $L(v)$, then check if $L(v) = L(\mathcal{S})$, which takes time $O(n \log n)$. If
$L(v) = L(\mathcal{S})$, then vertex $v$ satisfies the conditions of Reduction Rule 1,
otherwise, $L(v) \subsetneqq L(\mathcal{S})$, and we have to apply Stage-3: check if
$L(v) = L(C) \cap L(\mathcal{S})$, where $C$ is the connected component of $F_i'$ that
contains $v$. Stage-3 also takes time $O(n \log n)$. If $L(v) = L(C) \cap L(\mathcal{S})$,
then vertex $v$ satisfies the conditions of Reduction Rule 1, otherwise, no. In summary,
it takes time $O(n^2)$ to decide whether a vertex $v$ in $F_i'$ satisfies the conditions
of Reduction Rule 1, relative to $F_j'$.

Since there are $O(n)$ vertices in $F_i'$, it takes time $O(n^3)$ to decide whether
Reduction Rule 1 is applicable on some vertex in $F_i'$, relative to $F_j'$. For the
instance $(F_1', \ldots, F_m'; k')$, there are $O(m^2)$ pairs of $X$-forests, hence,
it takes time $O(m^2 n^3)$ to decide whether Reduction Rule 1 is applicable on
the instance. Since there are $O(mn)$ edges in the instance, applying Reduction
Rule 1 on $(F_1', \ldots, F_m'; k')$ until it is unapplicable takes time $O(m^3 n^4)$.

For two $X$-forests $F_i'$ and $F_j'$, $i < j$, and a connected component $C$ of
$F_i'$ (assuming the first label in $L(C)$ is $\alpha$), it takes time $O(n)$ to find
the connected component $C'$ of $F_j'$ that contains $\alpha$, and time $O(n)$
to check if $L(C) = L(C')$. Thus, deciding if $F_i'$ and $F_j'$ satisfy the label-set
isomorphism property takes time $O(n^2)$, and deciding if the instance
$(F_1', \ldots, F_m'; k')$ satisfies the label-set isomorphism property takes
time $O(mn^2)$.

For the $X$-forest $F_1'$, and an MSS $S$ in $F_1'$, it takes time $O(mn)$ to
check whether $S$ is an MSS in all $X$-forest in the instance. Thus, applying
Reduction Rule 2 on $(F_1', \ldots, F_m'; k')$ until it is unapplicable takes time
$O(mn^2)$.

In summary, between any two consecutive branches, the computational path
$\mathcal{P}$ takes time $O(m^3 n^4)$. Combining this with the following
easily-verified facts: (1) checking whether there is an $X$-forest $F$ in the
instance that has no MSS takes time $O(mn)$; (2) deciding whether an MSS
satisfies the given condition of one of the cases takes time $O(mn^3)$ (in
particular, deciding whether the instance satisfies the condition of Case 3.2 or
3.3 takes time $O(mn^3)$); (3) applying each branching rule takes time $O(mn)$;
and (4) the computational path $\mathcal{P}$ contains at most $k$ branches, we
conclude that the time complexity of the algorithm Alg-{\sc hMaf} is $O(2.42^k m^3n^4)$.
\end{proof}
\end{theorem}

\section{Parameterized Algorithm for the {\sc sMaf} Problem}

In this section, we present a parameterized algorithm for the {\sc sMaf} problem.
Remark that the {\sc sMaf} problem is much more complicated than the {\sc hMaf} problem, because of the flexibility about the binary resolutions of an $X$-forest. For example, given an instance $(F_1,F_2,\ldots,F_m;k)$ of the {\sc Maf} problem, and two labels $a$ and $b$ that are siblings, but are not a sibling-pair in $F_1$. Let $F^*$ be an arbitrary MAF for the $X$-forests in $(F_1,F_2,\ldots,F_m;k)$. If $(F_1,F_2,\ldots,F_m;k)$ is an instance of the {\sc hMaf} problem, then there are just three cases for labels $a$ and $b$ in $F^*$: label $a$ is a single-vertex tree; label $b$ is a single-vertex tree; labels $a$ and $b$ are in the same connected component in $F^*$, which implies that $a$ and $b$ are siblings in $F^*$. However, if $(F_1,F_2,\ldots,F_m;k)$ is an instance of the {\sc sMaf} problem, we cannot get a similar conclusion, since even though $a$ and $b$ are in the same connected component in $F^*$, they may not be siblings in $F^*$ (because there may not exist a binary resolution $F_1^{B}$ of $F_1$ such that $a$ and $b$ are siblings in $F_1^{B}$, and $F^*$ is a subforest of $F_1^{B}$). Thus, some branching rules for the {\sc hMaf} problem (in particular, the branching rules for Case 3) are not feasible for the {\sc sMaf} problem. But fortunately, if labels $a$ and $b$ are a sibling-pair in $F_1$, even though $(F_1,F_2,\ldots,F_m;k)$ is an instance of the {\sc sMaf} problem, we also have the three cases for $a$ and $b$ in $F^*$.

In the remaining parts of this section, we firstly present the $m$-BR*-process, which is an extension of the $m$-BR-process to the soft version, then analyze the detailed branching rules for the minimum MSS $S$ of the instance, according to the size of $S$.

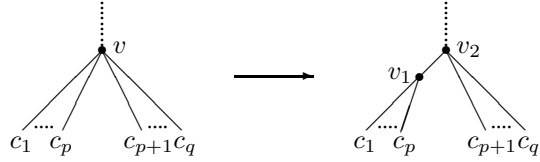
\begin{figure}[h]
\begin{center}
\begin{picture}(300,35)
\put(50,-45){\begin{picture}(0,0)
 \put(30,60){\circle*{3}}
 \multiput(30,60)(0,2){10}{\circle*{1}}

 \put(30,60){\line(-1,-1){30}}
 \put(30,60){\line(-1,-2){15}}
 \put(30,60){\line(1,-1){30}}
 \put(30,60){\line(1,-2){15}}

 \multiput(5,31)(2,0){4}{\circle*{1}}
 \multiput(48,31)(2,0){4}{\circle*{1}}

 \put(34,59){\small $v$}
 \put(-5,23){\small $c_1$}
 \put(10,23){\small $c_p$}
 \put(38,23){\small $c_{p+1}$}
 \put(57,23){\small $c_q$}
\end{picture}}

 \put(130,5){\vector(1,0){30}}

\put(180,-45){\begin{picture}(0,0)
 \put(30,60){\circle*{3}}
 \multiput(30,60)(0,2){10}{\circle*{1}}

 \put(30,60){\line(-1,-1){10}}
 \put(30,60){\line(1,-1){30}}
 \put(30,60){\line(1,-2){15}}

 \put(20,50){\circle*{3}}
 \put(20,50){\line(-1,-1){20}}
 \put(20,50){\line(-1,-3){7}}

 \multiput(5,31)(2,0){4}{\circle*{1}}
 \multiput(48,31)(2,0){4}{\circle*{1}}

 \put(34,59){\small $v_2$}
 \put(8,49){\small $v_1$}
 \put(-5,23){\small $c_1$}
 \put(10,23){\small $c_p$}
 \put(38,23){\small $c_{p+1}$}
 \put(57,23){\small $c_q$}
\end{picture}}

\end{picture}

\end{center}
\vspace*{5mm}
\caption{An illustration of the expansion for the children subset $\{c_1,\ldots,c_p\}$ of $v$.}
\label{fig:expand}
\end{figure}

First of all, we give some related definitions, which follows the ones given in \cite{17}. Given an $X$-forest $F$ and a vertex $v$ in $F$ with a children set $\{c_1,\ldots,c_p,c_{p+1},\ldots,c_q\}$ ($2 \leq p < q$). The {\it expansion} for the children subset $\{c_1,\ldots,c_p\}$ of $v$ (or expanding the children subset $\{c_1,\ldots,c_p\}$ of $v$), is defined as splitting the vertex $v$ into two vertices $v_1$ and $v_2$ such that $v_1$ is the child of $v_2$, and dividing the children of $v$ into two subsets $\{c_1,\ldots,c_p\}$ and $\{c_{p+1},\ldots,c_q\}$ that become the children-sets of $v_1$ and $v_2$ respectively. Figure~\ref{fig:expand} gives an illustration of the expansion. The edge between $v_1$ and $v_2$ is the {\it expanding edge} of the subset $\{c_1,\ldots,c_p\}$.

\subsection{The $m$-BR*-process}

In this subsection, we firstly give two rules -- Reduction Rule 1* and Branching Rule 1*, which are the extensions of Reduction Rule 1 and Branching Rule 1 to the soft version, respectively. Then based on the two rules, we extend the $m$-BR-process to the $m$-BR*-process. Let $(F_1,F_2,\ldots,F_m;k)$ be an instance of the {\sc sMaf} problem.

\smallskip

\noindent {\bf Reduction Rule 1*.} Let $\mathcal{C}_{F_i} = \{C_1,\ldots,C_t\}$ ($t \geq 1$) be a subset of the connected components in $X$-forest $F_i$, $1 \leq i \leq m$.

(1). If there exists a vertex $v$ in the connected component $C$ of $X$-forest $F_j$, $j \neq i$, such that $L(v) = L(C) \cap (L(C_1) \cup \ldots \cup L(C_t))$, then remove the edge $e$ between $v$ and $v$'s parent (if one exists) in $F_j$.

(2). If there exists a vertex $v$ in the connected component $C$ of $X$-forest $F_j$, $j \neq i$, with children set $\{c_1,\ldots,c_p,c_{p+1},\ldots,c_q\}$ ($2 \leq p < q$) such that $L(c_1) \cup \ldots \cup L(c_p) = L(C) \cap (L(C_1) \cup \ldots \cup L(C_t))$, then expand the set $\{c_1,\ldots,c_p\}$ in $F_j$ and remove the expanding edge $e$.

\smallskip

In the following, we give a critical lemma that is similar to Lemma 3 in~\cite{17}.

\begin{lemma}
\label{lem:expand}
Let $(F_1,F_2,\ldots,F_m;k)$ be an instance of the {\sc sMaf} problem, and $F^*$ be an MAF for the $X$-forests in $(F_1,F_2,\ldots,F_m;k)$. Let $v$ be a vertex in $F_i$ ($1 \leq i \leq m$) with a children set $\{c_1,\ldots,c_p,c_{p+1},\ldots,c_q\}$ ($2 \leq p < q$), and $F'_i$ be the forest obtained by expanding $\{c_{1},\ldots,v_p\}$ in $F_i$. If for any label $l \in L(c_1) \cup \ldots \cup L(c_p)$ and any label $l' \in L(c_{p+1}) \cup \ldots \cup L(c_q)$, there is no path between $l$ and $l'$ in $F^*$, then $F^*$ is also an MAF for the $X$-forests in $(F_1,\ldots,F'_i,\ldots,F_m;k)$.

\begin{proof}
Suppose that $F^*$ is not an MAF for the $X$-forests in $(F_1,\ldots,F'_i,\ldots,F_m;k)$. Then we have that there does not exist a binary resolution $F'^B_i$ of $F'_i$ such that $F^*$ is a subforest of $F^B_i$. It is easy to see that the difference between $F_i$ and $F'_i$ is the expansion of $\{c_{1},\ldots,c_p\}$. Thus, if there does not exist such a binary resolution $F'^B_i$ of $F'_i$,
then there exists a connected component in $F^*$ that contains labels $l_1 \in L(c_1) \cup \ldots \cup L(c_p)$ and $l_2 \in L(c_{p+1}) \cup \ldots \cup L(c_q)$, contradicting the fact that for any label $l \in L(c_1) \cup \ldots \cup L(c_p)$ and any label $l' \in L(c_{p+1}) \cup \ldots \cup L(c_q)$, there is no path between $l$ and $l'$ in $F^*$. Thus, the supposition is incorrect, and $F^*$ is also an MAF for the $X$-forests in $(F_1,\ldots,F'_i,\ldots,F_m;k)$.
\end{proof}
\end{lemma}

For the situation of Reduction Rule 1*, we say that Reduction Rule 1* is {\it applicable on $F_j$ relative to $F_i$}.
Let $(F_1,\ldots,F'_j,\ldots,F_m;k')$ be the instance obtained by applying Reduction Rule 1* on $(F_1,F_2,\ldots,F_m;k)$ with edge $e$ removed from $F_j$ (or with the expanding edge $e$ removed from $F_{j}^{E}$, where $F_{j}^{E}$ is the $X$-forest obtained by expanding the set $\{c_1,\ldots,c_p\}$ in $F_j$). Similar to the analysis for Reduction Rule 1, if $Ord(F_j) = Ord_{max}(F_1,F_2,\ldots,F_m;k)$, then $k' = k-1$, otherwise, $k' = k$. By Lemma~\ref{lem:expand}, we can easily get the following lemma.

\begin{lemma}
\label{lem:Reduction Rule 1*}
Instances $(F_1,F_2,\ldots,F_m;k)$ and $(F_1,\ldots,F'_j,\ldots,F_m;k')$ have the same collection of solutions.
\end{lemma}

Branching Rule 1* for $(F_1,F_2,\ldots,F_m;k)$ is presented as follows. Note that Reduction Rule 1* is also assumed unapplicable on $(F_1,F_2,\ldots,F_m;k)$.

\smallskip

\noindent {\bf Case 1*.} For a connected component $C$ in $F_i$, $1 \leq i \leq m$, there exists a vertex $v$ with a children set $\{c_1,\ldots,c_p,c_{p+1},\ldots,c_q\}$ ($1 \leq p < q$) in $F_j$, $j \neq i$, such that $(L(c_1) \cup \ldots \cup L(c_p))  \subseteq L(C)$, and $(L(c_{p+1}) \cup \ldots \cup L(c_q)) \cap L(C) = \emptyset$.

\smallskip

\noindent {\bf Branching Rule 1*.} Branch into two ways: [1] if $p=1$, then remove the edge between $v$ and $c_1$ from $F_j$, otherwise, expand the set $\{c_1,\ldots,c_p\}$ in $F_j$, and remove the expanding edge; [2] if $p+1 = q$, then remove the edge between $v$ and $c_q$ from $F_j$, otherwise, expand the set $\{c_{p+1},\ldots,c_q\}$ in $F_j$, and remove the expanding edge.

\begin{lemma}
\label{lem:Branching Rule 1*}
Branching Rule 1* is safe.
\end{lemma}

According to Reduction Rule 1* and Branching Rule 1*, the $m$-BR*-process can be defined, analogously to the $m$-BR-process. Note that for each edge removed by Branching Rule 1*, there exists a connected label-pair for it. Thus, the related lemmata and theorems for the $m$-BR-process are also feasible for the $m$-BR*-process.

\subsection{Analysis for Maximal Sibling Set of {\sc sMaf}}

In the following discussion, we assume that the instance $(F_1,F_2,\ldots,F_m;k)$ satisfies the label-set isomorphism property, i.e., Reduction Rule 1* and Branching Rule 1* are unapplicable on $(F_1,F_2,\ldots,F_m;k)$.

\smallskip

\noindent{\bf Reduction Rule 2*.} If there exist two labels $a$ and $b$ that are siblings in all $X$-forests, then {\it group} $a$ and $b$ into an un-decomposable structure, and mark the unit with the same label in all $X$-forests.

\smallskip

To implement Reduction Rule 2*, if the common parent of $a$ and $b$ in $F_i$, $1 \leq i \leq m$, has no other child, then we simply remove labels $a$ and $b$ and label the common parent of with $\overline{ab}$, otherwise, we remove label $a$ and relabel the leaf $b$ with new label $\overline{ab}$. In the further processing of $F_1,F_2,\ldots$, and $F_m$, we can treat $\overline{ab}$ as a new leaf in the forests. This step also replaces the label-set $X$ with a new label-set $(X \setminus \{a,b\}) \cup \{\overline{ab}\}$.

\begin{lemma}
\label{lem:group of two siblings}
For any instance $(F_1,F_2,\ldots,F_m;k)$ of the {\sc sMaf} problem, if $a$ and $b$ are siblings in all $X$-forests in it, then there exists an MAF for the $X$-forests in it, in which $a$ and $b$ are a sibling pair.

\begin{proof}
Let $F^*$ be an MAF for the $X$-forests in $(F_1,F_2,\ldots,F_m;k)$. If $a$ and $b$ are a sibling-pair in $F^*$, then the lemma holds true. Thus in the following discussion, we assume that $a$ and $b$ are not a sibling-pair in $F^*$.

If both $a$ and $b$ are single-vertex trees $F^*$, then by attaching single-vertex tree $a$ to single-vertex tree $b$ such that $a$ and $b$ are a sibling-pair, an agreement forest for the $X$-forests in $(F_1,F_2,\ldots,F_m;k)$ with a smaller order than $F^*$ can be constructed, contradicting the fact that $F^*$ is an MAF for the $X$-forests in $(F_1,F_2,\ldots,F_m;k)$.

If one of $a$ and $b$ is a single-vertex tree in $F^*$ (assume that $a$ is a single-vertex tree in $F^*$), then by attaching the single-vertex tree $a$ to the middle vertex of the edge incident to $b$ such that $a$ and $b$ are a sibling-pair, an agreement forest for the $X$-forests in $(F_1,F_2,\ldots,F_m;k)$ with a smaller order than $F^*$ can be constructed, contradicting the fact that $F^*$ is an MAF for the $X$-forests in $(F_1,F_2,\ldots,F_m;k)$.

By above discussion, we have that neither $a$ nor $b$ is a single-vertex tree in $F^*$. By removing the edge incident to $b$, and attaching the single-vertex tree $b$ to the middle vertex of the edge incident to $a$ such that $a$ and $b$ are a sibling-pair, another MAF for the $X$-forests in $(F_1,F_2,\ldots,F_m;k)$ can be constructed.
\end{proof}
\end{lemma}

\begin{lemma}
\label{lem:Reduction Rule 2*}
Let $(F'_1,F'_2,\ldots,F'_m;k)$ be the instance that is obtained by Reduction Rule 2* on $(F_1,F_2,\ldots,F_m;k)$ with grouping labels $a$ and $b$. Then every MAF for the $X$-forests in $(F'_1,F'_2,\ldots,F'_m;k)$ is also an MAF for the $X$-forests in $(F_1,F_2,\ldots,F_m;k)$.

\begin{proof}
By Lemma~\ref{lem:group of two siblings}, for each MAF $F'$ for the $X$-forests in $(F'_1,F'_2,\ldots,F'_m;k)$, we can easily get a corresponding MAF $F$ for the $X$-forests in $(F_1,F_2,\ldots,F_m;k)$ by expanding the grouped label $\overline{ab}$. Thus, we simply say that every MAF for the $X$-forests in $(F'_1,F'_2,\ldots,F'_m;k)$ is also an MAF for the $X$-forests in $(F_1,F_2,\ldots,F_m;k)$.
\end{proof}
\end{lemma}

In the following discussion, we assume that Reduction Rule 2* is unapplicable on $(F_1,F_2,\ldots,F_m;k)$. By Lemma~\ref{lem:no-MSS}, w.l.o.g., we assume that $F_1$ has a minimum MSS $S$ in $(F_1,\ldots,F_i,\ldots,F_m;k)$. Note that since $(F_1,F_2,\ldots,F_m;k)$ satisfies the label-set isomorphism property, the labels of $S$ are in the same connected component of $F_i$ for all $1 \leq i \leq m$.

\subsection{Case 2*: $|S|=2$}

\begin{figure}[h]
\begin{center}
\begin{picture}(300,80)

\put(0,-5){\begin{picture}(0,0)
 \put(30,70){\circle*{3}}
 \put(30,70){\line(-1,-1){40}}
 \put(30,70){\line(-1,-2){6}}
 \put(30,70){\line(1,-2){6}}
 \put(16,56){\circle*{3}}
 \put(16,56){\line(1,-1){10}}
 \put(5,45){\circle*{3}}
 \put(5,45){\line(1,-1){10}}
 \put(5,45){\line(0,-1){10}}

 \put(30,70){\line(1,-1){40}}
 \put(50,50){\circle*{3}}
 \put(50,50){\line(-1,-1){10}}
 \put(50,50){\line(0,-1){10}}

 \put(-10,30){\circle*{3}}
 \put(-14,20){\small $a$}
 \put(70,30){\circle*{3}}
 \put(68,20){\small $b$}

  \put(24,15){\small $F$}
\end{picture}}

\put(80,0){\small (1)}

\put(70,35){\vector(1,0){30}}

\put(110,-5){\begin{picture}(0,0)
 \put(30,70){\circle*{3}}
 \put(30,70){\line(-1,-1){40}}
 \put(30,70){\line(-1,-2){6}}
 \put(30,70){\line(1,-2){6}}
 \put(16,56){\circle*{3}}
 \thicklines
 \put(16,56){\line(1,-1){10}}
 \thinlines
 \put(5,45){\circle*{3}}
 \thicklines
 \put(5,45){\line(1,-1){10}}
 \thinlines
 \put(15,35){\circle*{3}}
 \put(15,35){\line(1,-1){10}}
 \put(15,35){\line(0,-1){10}}

 \put(30,70){\line(1,-1){40}}
 \put(50,50){\circle*{3}}
  \thicklines
 \put(50,50){\line(-1,-1){10}}
  \thinlines
 \put(40,40){\circle*{3}}
 \put(40,40){\line(-1,-1){10}}
 \put(40,40){\line(0,-1){10}}

 \put(-10,30){\circle*{3}}
 \put(-14,20){\small $a$}
 \put(70,30){\circle*{3}}
 \put(68,20){\small $b$}

 \put(24,13){\small $F^e$}
\end{picture}}

\put(200,-5){\begin{picture}(0,0)
 \put(30,70){\circle*{3}}
 \put(13,68){\small $p_{x_1}$}
 \put(30,40){\circle*{3}}
 \put(15,40){\circle*{3}}
 \put(45,40){\circle*{3}}
 \multiput(30,70)(0,-2){15}{\circle*{1}}
 \put(30,70){\line(-1,-2){15}}
 \multiput(30,70)(1,-2){15}{\circle*{1}}
 \put(10,30){\small $x_1$}
 \put(26,30){\small $x_{i_1}$}
 \put(40,30){\small $x_{i_z}$}

 \put(10,27){$\underbrace{\rule{1.4cm}{0cm}}$}
 \put(24,10){\small $S_1$}

 \put(30,70){\line(2,-1){60}}
 \put(90,40){\circle*{3}}

 \put(90,40){\line(-1,-2){7}}
 \put(90,40){\line(1,-2){7}}
 \put(83,26){\line(1,0){14}}

 \put(99,35){\small $\ldots$}

 \multiput(30,70)(3,-1){30}{\circle*{1}}
 \put(120,40){\circle*{3}}

 \put(120,40){\line(-1,-2){7}}
 \put(120,40){\line(1,-2){7}}
 \put(113,26){\line(1,0){14}}

 \put(86,42){$\overbrace{\rule{1.3cm}{0cm}}$}
 \put(100,51){\small $S_2$}

 \put(60,5){\small (2)}
\end{picture}}

\end{picture}

\end{center}
\vspace*{-5mm}
\caption{(1). An illustration of $X$-forest $F$ and its corresponding $F^e$. The edge-set $E_{F^e}(a,b)$ consists of the edges in $F^e$ that are in bold. (2). An illustration of Case 3*. The triangles are subtrees.}
\label{fig:|S|>=3}
\end{figure}
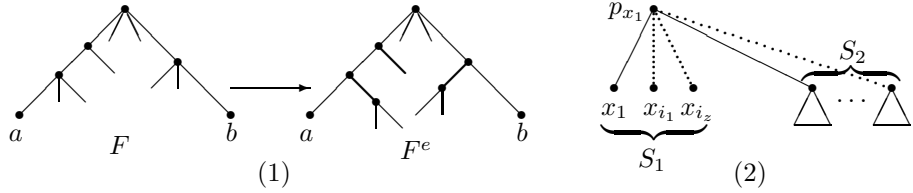

In this subsection, we assume that $S = \{a,b\}$.
Given an $X$-forest $F$ in which labels $a$ and $b$ are in the same connected component, denote by $Chd(v)$ the set containing all children of $v$, for any vertex $v$ in $F$, denote by $\mathcal{P}_{F}(a,b)$ the set containing all internal vertices in the path connecting $a$ and $b$ in $F$, except vertex $LCA_{F}(a,b)$, and denote by $F^e$ the $X$-forest obtained by expanding set $Chd(v) \setminus (\mathcal{P}_{F}(a,b) \cup \{a,b\})$, for all $v \in \mathcal{P}_{F}(a,b)$ such that $|Chd(v) \setminus (\mathcal{P}_{F}(a,b) \cup \{a,b\})| \geq 2$. See Figure~\ref{fig:|S|>=3} (1) for an illustration. It is easy to see that each vertex in $\mathcal{P}_{F^e}(a,b)$ has degree 3 in $F^e$.
Denote by $E_{F^e}(a,b)$ the edge-set containing all edges in $F^e$ that are incident to the vertices in $\mathcal{P}_{F^e}(a,b)$, but are not on the path connecting $a$ and $b$ in $F^e$. Obviously, all expanding edges are in $E_{F^e}(a,b)$, and $|\mathcal{P}_{F}(a,b)| = |E_{F^e}(a,b)|$. Note that $E_{F^e}(a,b)$ does not contain the edges incident to $LCA_{F}(a,b)$.

\smallskip

\noindent{\bf Case 2.1*.} There exists an $X$-forest $F_p$ such that $|\mathcal{P}_{F_p}(a,b)| \geq 2$.

\smallskip

\noindent{\bf Branching Rule 2.1*.} Branch into three ways: [1] remove the edge incident to $a$ in all $X$-forests; [2] remove the edge incident to $b$ in all $X$-forests; [3] construct the instance $(F_1,F_2^e,\ldots,F_m^e;k)$ by the expansion operation, and remove the edges in $E_{F^e_i}(a,b)$ for all $2 \leq i \leq m$.

\begin{lemma}
\label{lem:Branching Rule 2.1*}
Branching Rule 2.1* is safe, and satisfies the recurrence relation $T(k) \leq 2T(k-1) + T(k-2)$.

\begin{proof}
Let $F^*$ be an MAF for the $X$-forests in $(F_1,F_2,\ldots,F_m;k)$. Since $a$ and $b$ are a sibling-pair in $F_1$, labels $a$ and $b$ are a sibling-pair in any binary resolution of $F_1$. Thus, there are three possible cases for $a$ and $b$ in $F^*$.

(1). Label $a$ is a single-vertex tree in $F^*$. Thus, the first branch of Branching Rule 2.1* is correct. Since $Ord(F_1) = \ldots = Ord(F_m)$ and each $X$-forest in the new instance obtained by the first branch of Branching Rule 2.1* has order $Ord(F_1) + 1$, the value of the parameter in the new instance is $k - 1$.

(2). Label $b$ is a single-vertex tree in $F^*$. Similarly to the analysis for case (1), the second branch of Branching Rule 2.1* is correct, and the value of the parameter in the new instance is $k-1$.

(3). Labels $a$ and $b$ are a sibling-pair in $F^*$. Assume that there exists a vertex $v \in \mathcal{P}_{F_i}(a,b)$ (for some $2 \leq i \leq m$) such that some label $l_1$ of $L(Chd(v) \setminus (\mathcal{P}_{F_i}(a,b) \cup \{a,b\}))$ is in the same connected component of $F^*$ with some label of $X \setminus (L(Chd(v) \setminus (\mathcal{P}_{F_i}(a,b) \cup \{a,b\})))$, then we can get that $l_1$ is in the same connected component with $a$ and $b$ in $F^*$, which implies that $a$ and $b$ cannot be siblings in $F^*$. Thus, the assumption is incorrect. By Lemma~\ref{lem:expand}, the expansion operation for the set $Chd(v) \setminus (\mathcal{P}_{F_i}(a,b) \cup \{a,b\})$ for all $v \in \mathcal{P}_{F_i}(a,b)$ such that $|Chd(v) \setminus (\mathcal{P}_{F_i}(a,b) \cup \{a,b\})| \geq 2$ is correct, and the edges in $E_{F^e_i}(a,b)$ could be removed.

Without loss of generality, assume that $|E_{F^e_p}(a,b)| = \max \{|E_{F^e_2}(a,b)|,\ldots,|E_{F^e_m}(a,b)|\}$.
Since $E_{F^e_p}(a,b)$ is an ee-set of $F^e_p$, $|E_{F^e_p}(a,b)| \geq 2$, and $Ord(F^e_p) = Ord(F_p)$, we have that $Ord(F^e_p \setminus E_{F^e_p}(a,b)) \geq Ord(F_p)+2$ and $k' \leq k -2$, where $k'$ is the parameter of the new instance $(F_1,F^e_2 \setminus E_{F^e_2}(a,b),\ldots,F^e_m \setminus E_{F^e_m}(a,b);k')$.

Summarizing above discussion, the recurrence relation of Branching Rule 2.1* is $T(k) \leq 2T(k-1) + T(k-2)$.
\end{proof}
\end{lemma}

\noindent{\bf Case 2.2*.} For all $2 \leq i \leq m$, $|\mathcal{P}_{F_i}(a,b)| \leq 1$.

\smallskip

\noindent{\bf Case 2.2.1*.} There exists two $X$-forests $F_s$ and $F_t$ such that $|\mathcal{P}_{F_s}(a,b)| = |\mathcal{P}_{F_t}(a,b)| = 1$ and $L(\mathcal{P}_{F_s}(a,b)) \setminus \{a,b\} \neq L(\mathcal{P}_{F_t}(a,b)) \setminus \{a,b\}$.

\smallskip

\noindent{\bf Branching Rule 2.2.1*.} Branch into three ways: [1] remove the edge incident to $a$ in all $X$-forests; [2] remove the edge incident to $b$ in all $X$-forests; [3] construct the instance $(F_1,F_2^e,\ldots,F_m^e;k)$ by expansion operation, remove the edges in $E_{F^e_i}(a,b)$ for all $2 \leq i \leq m$, and apply $m$-BR*-process.

\begin{lemma}
\label{lem:Branching Rule 2.2.1*}
Branching Rule 2.2.1* is safe, and satisfies the recurrence relation $T(k) \leq 2T(k-1) + T(k-2)$.

\begin{proof}
Let $F^*$ be an MAF for the $X$-forests in $(F_1,F_2,\ldots,F_m;k)$. There are three possible cases for $a$ and $b$ in $F^*$.

(1-2). Label $a$ or $b$ is a single-vertex tree in $F^*$. Using the analysis for the first two cases in the proof of Lemma~\ref{lem:Branching Rule 2.1*}, we can also get that the first two branches of Branching Rule 2.2.1* are correct, and the two branches construct two new instances whose parameter values are $k-1$.

(3). Labels $a$ and $b$ are a sibling-pair in $F^*$. Using the analysis for the third case in the proof of Lemma~\ref{lem:Branching Rule 2.1*}, we can get that the third branch of Branching Rule 2.2.1* is also correct.
For the new instance $I = (F_1,F^e_2 \setminus E_{F^e_{2}}(a,b),\ldots,F^e_m \setminus E_{F^e_{m}}(a,b);k')$ obtained by the third branch, we can see that $I$ satisfies the 2-edge distance property and
that $k' = k-1$. The discussion about the $m$-BR*-process on $I$ is divided into two subcases.

(3.1) Branching Rule 1* is not applied during the $m$-BR*-process on $I$. By
Theorem~\ref{theo:special m-BR-process 0}, only one instance is obtained by the
$m$-BR*-process, whose parameter $k''$ has value not larger than $k'-1 = k-2$. Thus,
in this subcase, the recurrence relation of Branching Rule 2.2.1* is
$T(k) = 2T(k-1) + T(k'') \leq 2T(k-1) + T(k-2)$.

(3.2) Branching Rule 1* is applied during the $m$-BR*-process on $I$. By
Theorem~\ref{theo:special m-BR-process 1},
$T(k') \leq |\mathcal{C}_r| \cdot T(k' - r-1) + \ldots + |\mathcal{C}_h| \cdot T(k' - h-1)$,
where $k' = k-1$ and $\frac{|\mathcal{C}_r|}{2^r} + \ldots + \frac{|\mathcal{C}_h|}{2^h} = 1$.
Thus, in this subcase, the recurrence relation of Branching Rule 2.2.1* is
\[ T(k) \leq 2T(k-1) + |\mathcal{C}_r| \cdot T(k - r-2) + \ldots + |\mathcal{C}_h| \cdot T(k - h-2).\]
The characteristic polynomial of the above recurrence relation is
$p(x) = x^{h+2} - 2 x^{h+1} - |\mathcal{C}_r| \cdot x^{h-r} - \ldots - |\mathcal{C}_h|$.
Since $p(2) < 0$ and $p(1+\sqrt{2}) > 0$, the unique positive root of $p(x)$ has its value
bounded by $1+\sqrt{2}$. Therefore, if Branching Rule 1* is applied during the $m$-BR*-process
on $I$, then Branching Rule 2.2.1* satisfies the recurrence relation $T(k) \leq 2T(k-1) + T(k-2)$,
whose characteristic polynomial has its unique positive root $1+\sqrt{2}$.

Summarizing these discussions, we conclude that the recurrence relation of Branching
Rule 2.2.1* satisfies $T(k) \leq 2T(k-1) + T(k-2)$.
\end{proof}
\end{lemma}

\noindent{\bf Case 2.2.2*.} For any two $X$-forests $F_s$ and $F_t$ such that $|\mathcal{P}_{F_s}(a,b)| = |\mathcal{P}_{F_t}(a,b)| = 1$, $L(\mathcal{P}_{F_s}(a,b)) \setminus \{a,b\} = L(\mathcal{P}_{F_t}(a,b)) \setminus \{a,b\}$.

\smallskip

\noindent{\bf Case 2.2.2.1*.} For any $X$-forest $F_p$ such that $|\mathcal{P}_{F_p}(a,b)| = 1$, the unique vertex of $\mathcal{P}_{F_p}(a,b)$ is closer to $a$ than $b$ (or closer to $b$ than $a$).

\smallskip

\noindent{\bf Branching Rule 2.2.2.1*.} Branch into two ways: [1] remove the edge incident to $b$ in all $X$-forests if the unique vertex of $\mathcal{P}_{F_p}(a,b)$ is closer to $a$, otherwise, remove the edge incident to $a$ in all $X$-fores; [2] construct a new instance $(F_1,F_2^e,\ldots,F_m^e;k)$ by the expansion operation, and remove the edges in $E_{F^e_i}(a,b)$ for all $2 \leq i \leq m$.

\begin{lemma}
\label{lem:Branching Rule 2.2.2.1*}
Branching Rule 2.2.2.1* is safe, and satisfies the recurrence relation $T(k) \leq  2T(k-1)$.

\begin{proof}
Because of symmetry, we just analyze the subcase that the unique vertex of $P_{F_p}(a,b)$ is closer to $a$ than $b$, for all $X$-forest $F_p$ ($2 \leq p \leq m$) such that $|\mathcal{P}_{F_p}(a,b)| = 1$.

Let $F^*$ be an arbitrary MAF for the $X$-forests in $(F_1,F_2,\ldots,F_m;k)$. If $a$ and $b$ are in the same connected component in $F^*$, then $a$ and $b$ are a sibling-pair in $F^*$, and the second branch of Branching Rule 2.2.2.1* is correct. For the new instance $(F_1,F^e_2 \setminus E_{F^e_{2}}(a,b),\ldots,F^e_m \setminus E_{F^e_{m}}(a,b);k')$, we have that $k' = k -1$.

If $a$ and $b$ are not in the same connected component in $F^*$, then at least one of $a$ and $b$ is a single-vertex tree in $F^*$. If both $a$ and $b$ are single-vertex trees, then there exists some label of $L(P_{F_p}(a,b)) \setminus \{a\}$ that is in the same connected component $C^*$ with some label of $X \setminus L(P_{F_p}(a,b))$ in $F^*$, otherwise, an agreement forest with a smaller order than $F^*$ can be constructed by attaching single-vertex tree $a$ to single-vertex tree $b$ such that $a$ and $b$ are a sibling-pair, contradicting the fact that $F^*$ is an MAF for the $X$-forests in $(F_1,F_2,\ldots,F_m;k)$.

Let $v^*$ be the vertex in $C^*$ such that $L(v^*) = L(P_{F_p}(a,b)) \cap L(C^*)$. By removing the edge between $v^*$ and $v^*$'s parent, and attaching single-vertex tree $a$ to single-vertex tree $b$ such that $a$ and $b$ are a sibling-pair, another MAF  for the $X$-forests in $(F_1,F_2,\ldots,F_m;k)$ can be constructed, which implies that the second branch of Branching Rule 2.2.2.1* is correct.

If only label $b$ is a single-vertex tree, then the first branch of Branching Rule 2.2.2.1* is correct, and the value of the parameter in the new resulting instance is $k-1$. In the following, we show that if only label $a$ is a single-vertex tree in $F^*$, then there exists another MAF in which $a$ and $b$ are a sibling-pair, implying that the second branch of Branching Rule 2.2.2.1* is correct.

Note that if there does not exist any label of $L(\mathcal{P}_{F_p}(a,b)) \setminus \{a\}$ that is in the same connected component $C^*$ with any label $X \setminus L(\mathcal{P}_{F_p}(a,b))$ in $F^*$, then an agreement forest with a smaller order than $F^*$ can be constructed by attaching the single-vertex tree $a$ to the middle vertex of the edge incident to $b$ such that $a$ and $b$ are a sibling-pair. Thus, there exists some label of $L(\mathcal{P}_{F_p}(a,b)) \setminus \{a\}$ that is in the same connected component $C^*$ with some label of $X \setminus L(\mathcal{P}_{F_p}(a,b))$ in $F^*$. Let $v^*$ be the vertex in $C^*$ such that $L(v^*) = (L(\mathcal{P}_{F_p}(a,b)) \setminus \{a\}) \cap L(C^*)$. By removing the edge between $v^*$ and $v^*$'s parent, and attaching single-vertex tree $a$ to the middle vertex of the edge incident to $b$ such that $a$ and $b$ are a sibling-pair, another MAF for the $X$-forests in $(F_1,F_2,\ldots,F_m;k)$ can be constructed.

Summarizing above analysis, the recurrence relation of Branching Rule 2.2.2.1* is $T(k) \leq 2T(k-1)$.

Remark that for this case, we have proved above that if only label $a$ is a single-vertex tree in $F^*$, then there exists another MAF for the $X$-forests in $(F_1,F_2,\ldots,F_m;k)$, in which $a$ and $b$ are a sibling-pair. However, if only label $b$ is a single-vertex tree in $F^*$, then there may not exist an MAF in which $a$ and $b$ are a sibling-pair. We give a specific example as follows. Assume that $L(\mathcal{P}_{F_p}(a,b)) \setminus \{a\} = \{c,d\}$, and labels $a$, $c$, and $d$ have a common parent in $F$ for all $F$ in the instance such that $|\mathcal{P}_{F}(a,b)| = 1$. It is easy to see that $a$ is in the same connected component $C^*$ with some label of $\{c,d\}$. Thus, we can assume that $a$, $c$, and $d$ are in the same connected component in $F^*$, and the structure about $a$, $c$, and $d$ in $F^*$ is $(c,(a,d))$ (because we can assume that the structure of the subtree $T_{F_1}[\{a,c,d\}]$ is $(c,(a,d))$). For this situation, if we try to construct an agreement forest $F'$ by doing several simple operations on $F^*$ such that $a$ and $b$ are a sibling-pair in $F'$, then the two edges incident to $c$ and $d$ respectively should be removed from $F^*$, and the single-vertex tree $b$ should be attached to the middle vertex of the edge incident to $a$. Thus, we have that $Ord(F') = Ord(F^*)+1$, and $F'$ is not an MAF for the $X$-forests in $(F_1,F_2,\ldots,F_m;k)$.
\end{proof}
\end{lemma}

\noindent{\bf Case 2.2.2.2*.} There exists two $X$-forests $F_s$ and $F_t$ such that $|\mathcal{P}_{F_s}(a,b)| = |\mathcal{P}_{F_t}(a,b)| = 1$, the unique vertex in $P_{F_s}(a,b)$ is closer to $a$ than $b$, and the unique vertex in $P_{F_t}(a,b)$ is closer to $b$ than $a$.

\smallskip

\noindent{\bf Reduction Rule 2.2.2.2*.} Construct the instance $(F_1,F_2^e,\ldots,F_m^e;k)$ by expansion operation, and remove the edges in $E_{F^e_i}(a,b)$ for all $2 \leq i \leq m$.

\begin{lemma}
\label{lem:Reduction Rule 2.2.2.2*}
Reduction Rule 2.2.2.2* on an instance $I$ of the {\sc sMaf} problem produces an instance that is a yes-instance if and only if $I$ is a yes-instance.

\begin{proof}
Let $F^*$ be an arbitrary MAF for the $X$-forests in $(F_1,F_2,\ldots,F_m;k)$. If $a$ and $b$ are in the same connected component in $F^*$, then $a$ and $b$ are a sibling-pair in $F^*$, and Reduction Rule 2.2.2.2* is correct.

If $a$ and $b$ are not in the same connected component in $F^*$, then at least one of $a$ and $b$ are single-vertex trees. Moreover, we have that there exists a connected component $C^*$ in $F^*$ that contains some label of $L(P_{F_s}(a,b)) \setminus \{a\}$ and some label of $X \setminus (L(P_{F_s}(a,b)) \setminus \{a\})$, otherwise, an agreement forest with a smaller order than $F^*$ can be constructed, in which $a$ and $b$ are a sibling-pair.

In the following, we show that if $a$ and $b$ are not in the same connected component in $F^*$, then there exists another MAF for the $X$-forests in $(F_1,F_2,\ldots,F_m;k)$, in which $a$ and $b$ are a sibling-pair.

If both $a$ and $b$ are single-vertex trees in $F^*$, then by removing the edge between $v^*$ and $v^*$'s parent, where $v^*$ is the vertex in the connected component $C^*$ in  $F^*$ such that $L(v^*) = L(C^*) \cap (L(P_{F_s}(a,b)) \setminus \{a\})$, and attaching single-vertex tree $a$ to single-vertex tree $b$ such that $a$ and $b$ are a sibling-pair, another MAF for the $X$-forests in $(F_1,F_2,\ldots,F_m;k)$ can be constructed. Thus, Reduction Rule 2.2.2.2* is correct.

If only $b$ is a single-vertex tree in $F^*$, i.e., label $a$ is not a single-vertex tree in $F^*$, then $a$ is in the connected component $C^*$ in $F^*$. Since the unique vertex in $P_{F_t}(a,b)$ is closer to $b$ than $a$ in $F_t$, and $L(P_{F_s}(a,b)) \setminus \{a\} = L(P_{F_t}(a,b)) \setminus \{b\}$, we can get that there must exist a vertex $v^*$ in $C^*$ such that $L(v^*) = L(C^*) \cap (L(P_{F_s}(a,b)) \setminus \{a\})$ (label $a$ cannot be surrounded by the labels of $L(P_{F_s}(a,b)) \setminus \{a\}$, like the example we gave in the last paragraph of the proof for Lemma~\ref{lem:Branching Rule 2.2.2.1*}). For this situation, another MAF $F'$ can be constructed by removing the edge between $v^*$ and $v^*$'s parent, and attaching the single-vertex tree $b$ to the middle vertex of the edge incident to $a$ such that $a$ and $b$ are a sibling-pair, which implies that Reduction Rule 2.2.2.2* is correct.

Similar analysis is feasible for the case that only $a$ is a single-vertex tree in $F^*$.
\end{proof}
\end{lemma}

\subsection{Case 3* for $|S|\geq 3$}

Assume that $S = \{x_1,x_2,x_3,\ldots\}$. Since we assumed above that Reduction Rule 2* is unapplicable on $(F_1,F_2,\ldots,F_m;k)$, there exists an $X$-forest $F_p$, $2 \leq p \leq m$, such that there are two labels of $S$ that are not siblings in $F_p$.
Denote by $r_S$ the root of the connected component containing $S$ in $F_p$. W.l.o.g., we assume that the distance from $r_S$ to label $x_1 \in S$ is the largest one among the distances from $r_S$ to the labels of $S$. Since there are two labels of $S$ that are not siblings in $F_p$, we have that $L(p_{x_1}) \nsubseteq S$, where $p_{x_1}$ denotes the parent of $x_1$ in $F_p$, otherwise, $S$ is not the minimum MSS of $(F_1,F_2,\ldots,F_m;k)$.

Let $S_1 = Chd(p_{x_1}) \cap S$, where $Chd(p_{x_1})$ denotes the children set of $p_{x_1}$, and $S_2 = Chd(p_{x_1}) \setminus S_1$. See Figure~\ref{fig:|S|>=3}(2) for an illustration. Note again that $S_1 \subsetneqq S$.

\smallskip

\noindent{\bf Branching Rule 3*.} Branch into three ways: [1] if $|S_1| = 1$, remove the edge incident to the unique label in $S_1$ in all $X$-forests; otherwise, construct $F^E_p$ by expanding the set $S_1$ in $F_p$, remove the expanding edge $e_{S_1}$ in $F^E_p$, and apply $m$-BR*-process;
[2] if $|S_2| = 1$, remove the edge between between $p_{x_1}$ and the unique vertex in $S_2$ in $F_p$; otherwise, construct $F^E_p$ by expanding the set $S_2$ in $F_p$, and remove the expanding edge $e_{S_2}$ in $F^E_p$;
[3] if $|S \setminus S_1| = 1$, remove the edge incident to the label in $S \setminus S_1$ in all $X$-forests; otherwise, construct $F^E_1$ by expanding the set $S \setminus S_1$ in $F_1$, remove the expanding edge $e_{S \setminus S_1}$ in $F^E_1$, and apply $m$-BR*-process.

\begin{lemma}
\label{lem:Branching Rule 3*}
Branching Rule 3* is safe, and satisfies the recurrence relation $T(k) \leq 2T(k-1) + 2T(k-2)$.

\begin{proof}
Let $F^*$ be an arbitrary MAF for the $X$-forests in $(F_1,F_2,\ldots,F_m;k)$. We firstly consider the case that some label $l_1$ of $L(S_1)$ is in the same connected component $C^*$ with some label $l_2$ of $L(S_2)$ in $F^*$. By the structure of $F_p$, it is easy to see that there exists a vertex $v^*$ in $C^*$ such that $L(v^*) = L(C^*) \cap (L(S_1) \cup L(S_2))$. Assume that some label $l_3$ of $S \setminus S_1$ is in the same connected component with some label of $X \setminus (S \setminus S_1)$. Then by the structure of $F_1$, we can get that $l_1$, $l_2$, and $l_3$ are in the same connected component $C^*$. It is not hard to see that the subtree $T_{F_1}[S']$ is not isomorphic to the subtree $T_{F_p}[S']$, where $S' = \{l_1,l_2,l_3\}$. Thus, $F^*$ cannot be an agreement forest for the $X$-forests in $(F_1,F_2,\ldots,F_m;k)$, and the assumption is incorrect, i.e., any label of $S \setminus S_1$ cannot be in the same connected component with any label of $X \setminus (S \setminus S_1)$ in the case that some label $l_1$ of $L(S_1)$ is in the same connected component $C^*$ with some label $l_2$ of $L(S_2)$ in $F^*$. By Lemma~\ref{lem:expand}, the third branch of Branching Rule 3* is correct.

If any label of $L(S_1)$ is not in the same connected component with any label of $L(S_2)$ in $F^*$, then by the structure of $F_p$, we can get that either any label of $L(S_1)$ is not in the same connected component with any label of $X \setminus L(S_1)$ in $F^*$ or any label of $L(S_2)$ is not in the same connected component with any label of $X \setminus L(S_2)$ in $F^*$. Thus by Lemma~\ref{lem:expand}, either the first or the second branch of Branching Rule 3* is correct.

If $|S_1| \geq 2$ (note that $x_1 \in S_1$, thus we assume that $x_2$ is also in $S_1$), then there exists an $X$-forest $F_q$ in the instance such that $x_1$ and $x_2$ are not siblings, otherwise, Reduction Rule 2* is applicable. Thus, for the first branch of Branching Rule 3*, Branching Rule 1* is applied at least once during the $m$-BR*-process on $(F_1,\ldots, F^E_p \setminus \{e_{S_1}\},\ldots,F_m,k')$, where $k' = k-1$. By Theorem~\ref{theo:m-BR-process}, we have that $T(k') \leq |\mathcal{C}_r| \cdot T(k' - r) + \ldots + |\mathcal{C}_h| \cdot T(k' - h)$, where $\frac{|\mathcal{C}_r|}{2^r} + \ldots + \frac{|\mathcal{C}_h|}{2^h} = 1$ and $1 \leq r \leq h$.

Similarly, if $|S \setminus S_1| \geq 2$, then for the third branch of Branching Rule 3*, Branching Rule 1* is applied at least once during the $m$-BR*-process on $(F^E_1 \setminus \{e_{S \setminus S_1}\},\ldots, F_p,\ldots,F_m,k')$, and $T(k') \leq |\mathcal{C}_r| \cdot T(k' - r) + \ldots + |\mathcal{C}_h| \cdot T(k' - h)$.

Since $|S| \geq 3$, at least one of the two inequalities $|S_1| \geq 2$ and $|S \setminus S_1| \geq 2$ holds true. Therefore, the recurrence relation of Branching Rule 3* is 
$$T(k) \leq 2T(k-1) + |\mathcal{C}_r| \cdot T(k-1 - r) + \ldots + |\mathcal{C}_h| \cdot T(k-1 - h),$$ 
where $\frac{|\mathcal{C}_r|}{2^r} + \ldots + \frac{|\mathcal{C}_h|}{2^h} = 1$.

The characteristic polynomial of the above recurrence relation is
$p(x) = x^{h+1} - 2 x^{h} - |\mathcal{C}_r| \cdot x^{h-r} - \ldots - |\mathcal{C}_h|$.
Since $p(2) < 0$ and $p(1+\sqrt{3}) \geq 0$, the unique positive root of $p(x)$ has its value
bounded by $1+\sqrt{3}$. Therefore, Branching Rule 3* satisfies the recurrence relation $T(k) \leq 2T(k-1) + 2T(k-2)$, whose characteristic polynomial has its unique positive root $1+\sqrt{3}$.
\end{proof}
\end{lemma}

\begin{figure}[ht]
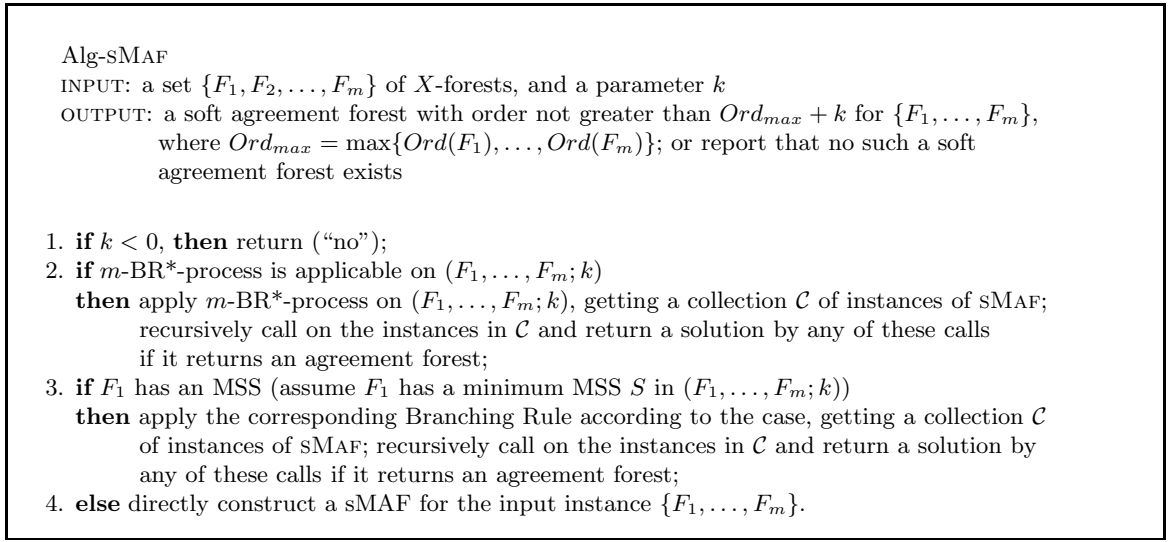

\vspace*{-3mm}
\setbox4=\vbox{\hsize18pc \noindent\strut
\begin{quote}
  \vspace*{-3mm} \footnotesize
  \hspace*{-8mm}
\hspace*{2mm}  Alg-{\sc sMaf}\\
  \hspace*{-5mm}
  {\sc input}: a set $\{F_1,F_2,\ldots,F_m\}$ of $X$-forests, and a parameter $k$\\
  \hspace*{-5mm}
  {\sc output}: a soft agreement forest with order not greater than $Ord_{max}+k$ for $\{F_1, \ldots, F_m\}$,\\
\hspace*{8mm} where $Ord_{max} = \max \{Ord(F_1), \ldots, Ord(F_m)\}$; or report that no such a  soft\\
\hspace*{8mm} agreement forest exists\\

\hspace*{-7mm}
  1.  {\bf if} $k < 0$, {\bf then} return (``no'');\\
  \hspace*{-7mm}
  2.  {\bf if} $m$-BR*-process is applicable on $(F_1, \ldots, F_m; k)$\\
  \hspace*{-3mm} {\bf then} apply $m$-BR*-process on $(F_1, \ldots, F_m; k)$, getting a collection
              $\mathcal{C}$ of instances of {\sc sMaf};\\
\hspace*{5.5mm}  recursively call on the instances in $\mathcal{C}$ and return a solution by any of these calls\\
\hspace*{5.5mm}   if it returns an agreement forest; \\
  \hspace*{-7mm}
  3.  {\bf if} $F_1$ has an MSS (assume $F_1$ has a minimum MSS $S$ in $(F_1, \ldots, F_m; k)$)\\
 \hspace*{-3mm} {\bf then} apply the corresponding Branching Rule according to the case, getting a collection $\mathcal{C}$\\
 \hspace*{5.5mm} of instances of {\sc sMaf}; recursively call on the instances in $\mathcal{C}$ and return a solution by\\
 \hspace*{5.5mm} any of these calls if it returns an agreement forest;\\
  \hspace*{-7mm}
  4. {\bf else} directly construct a sMAF for the input instance $\{F_1, \ldots, F_m\}$.

\end{quote} \vspace*{-6mm} \strut} $$\boxit{\box4}$$
\vspace*{-6mm}
\caption{A parameterized algorithm for the {\sc sMaf} problem}
\label{fig:alg-sMaf}
\vspace*{-2mm}
\end{figure}

Now we are ready to present the parameterized algorithm for the {\sc sMaf} problem, which is given in Figure~\ref{fig:alg-sMaf}.

\begin{theorem}
\label{theo:sMaf}
Algorithm Alg-{\sc sMaf} correctly solves the {\sc sMaf} problem in time $O(2.74^k m^3 n^5)$, where $n$ is the size of the label-set $X$ and $m$ is the number of $X$-forests in the input instance.

\begin{proof}
The proof for this theorem is similar to that for Theorem~\ref{theo:hMaf}. From the proof for Theorem~\ref{theo:hMaf}, we know that the time complexity for applying Reduction Rule 1 on the instance of the {\sc hMaf} problem until it is unapplicable,  contributes directly to the polynomial part of the time complexity of the algorithm Alg-{\sc hMaf}. Thus, in the following, we detailedly analyze the time complexity for applying Reduction Rule 1* on the instance $(F_1,F_2,\ldots,F_m;k)$ of the {\sc sMaf} problem until it is unapplicable. First of all, we also do some preparation work, as that given in the proof for Theorem~\ref{theo:hMaf}.

For simplicity of analysis, we firstly analyze the time complexity to decide whether Reduction Rule 1* is applicable on $F_i$ relative to $F_j$, $1 \leq i < j \leq m$. Since Reduction Rule 1*(1) is the same as Reduction Rule 1, it takes time $O(n^3)$ to decide whether Reduction Rule 1*(1) is applicable on $F_i$ relative to $F_j$. In the following, we analyze the time complexity to decide whether Reduction Rule 1*(2) is applicable on $F_i$ relative to $F_j$, under the assumption that Reduction Rule 1*(1) is unapplicable on $F_i$ relative to $F_j$.

Given an $X$-forest $F$, and a subset $X'$ of $X$, denote by $\mathcal{C}_{F}(X')$ the collection containing all connected components in $F$ that have some label in $X'$ (w.l.o.g., we assume that $\mathcal{C}_{F}(X')$ contains the serial numbers of the connected components in $F_j$, which are comparable).
In the following, we analyze the time to decide whether Reduction Rule 1*(2) is applicable on some children of vertex $v$ in $F_i$ relative to $F_j$. Our goal is to find a proper subset $V^*$ of the children set $Chd(v)$ of $v$ such that $L(V^*) = L(C_i) \cap (\cup_{C \in \mathcal{C}_{F_j}(L(V^*))}L(C))$, where $C_i$ is the connected component in $F_i$ containing vertex $v$. Initialize set $V'$ with $Chd(v)$.

Stage-1: For each connected component $C_j$ in $F_j$, check if $L(C_j) \cap L(C_i)$ is a subset of $L(v)$. If $L(C_j) \cap L(C_i)$ is not a subset of $L(v)$, then for any $c \in V'$ such that $L(c) \cap L(C_j) \neq \emptyset$, $V' = V' \setminus \{c\}$ (because $c$ cannot be in any $V^*$). If after Step-1, $|V'| \leq 1$, then Reduction Rule 1*(2) is unapplicable on the vertex $v$ in $F_i$ relative to $F_j$ (note that since we assumed that Reduction Rule 1*(1) is unapplicable, if $|V'| = 1$, then the edge between $v$ and the unique vertex in $V'$ could be removed by Reduction Rule 1*(1)). Since $F_j$ has at most $n$ connected components, this step takes time $O(n^2)$. Assume that $V' = \{c_1,\ldots,c_t\}$ ($2 \leq t \leq n$). We sort the elements in $\mathcal{C}_{F_j}(L(c))$ for each $c \in V'$, which takes time $O(n^2 \log n)$.

Step-2: Initialize $V_1 = \{c_1\}$; while there exists a connected component $c_s$ ($2 \leq s \leq t$) such that $c_s \notin V_1$ and $\mathcal{C}_{F_j}(L(c_s)) \cap \mathcal{C}_{F_j}(L(V_1)) \neq \emptyset$, include it into $V_1$. If $V_1$ is a proper subset of $Chd(v)$, then Reduction Rule 1*(2) is applicable on the set $V_1$, otherwise, applying the following step. Since $v$ has at most $n$ children, and $F_j$ has at most $n$ connected components, this step takes time $O(n^2)$.

Step-3: If $\mathcal{C}_{F_j}(L(c_2)) \cap \mathcal{C}_{F_j}(L(c_1)) = \emptyset$, then initialize $V_1 = \{c_2\}$, otherwise, initialize $V_1 = \emptyset$; while there exists a connected component $c_s$ ($3 \leq s \leq t$) such that $c_s \notin V_1$, $\mathcal{C}_{F_j}(L(c_s)) \cap \mathcal{C}_{F_j}(L(c_1)) = \emptyset$, and $\mathcal{C}_{F_j}(L(c_s)) \cap \mathcal{C}_{F_j}(L(V_1)) \neq \emptyset$, include it into $V_1$. If $V_1 \neq \emptyset$ and $V_1$ is a proper subset of $Chd(v)$, then Reduction Rule 1*(2) is applicable on the set $V_1$, otherwise, applying the following step. This step also takes time $O(n^2)$.

Step-$h+1$ (for all $3 \leq h \leq t$): If $\mathcal{C}_{F_j}(L(c_h)) \cap \mathcal{C}_{F_j}(L(V'')) = \emptyset$, where $V'' = \{v_1,\ldots,v_{h-1}\}$, then initialize $V_1 = \{c_h\}$, otherwise, initialize $V_1 = \emptyset$; while there exists a connected component $c_s$ ($h+1 \leq s \leq t$) such that $c_s \notin V_1$, $\mathcal{C}_{F_j}(L(c_s)) \cap \mathcal{C}_{F_j}(L(V'')) = \emptyset$, and $\mathcal{C}_{F_j}(L(c_s)) \cap \mathcal{C}_{F_j}(L(V_1)) \neq \emptyset$, include it into $V_1$. If $V_1 \neq \emptyset$ and $V_1$ is a proper subset of $Chd(v)$, then Reduction Rule 1*(2) is applicable on the set $V_1$, otherwise, applying the following feasible step. This step also takes time $O(n^2)$.

By above analysis, since $2 \leq t \leq n$, we can get that it takes time $O(n^3)$ to decide whether Reduction Rule 1*(2) is applicable on some children of vertex $v$ in $F_i$, relative to $F_j$. Since there are $O(n)$ vertices in $F_i$, it takes time $O(n^4)$ to decide whether Reduction Rule 1*(2) is applicable on $F_i$, relative to $F_j$. Combining the fact that it takes time $O(n^3)$ to decide whether Reduction Rule 1*(1) is applicable on $F_i$, relative to $F_j$, we have that it takes time $O(n^4)$ to decide whether Reduction Rule 1* is applicable on $F_i$, relative to $F_j$. For the instance $(F_1,F_2,\ldots,F_m;k)$, there are $O(m^2)$ pairs of $X$-forests in the instance, hence it takes time $O(m^2 n^4)$ to decide whether Reduction Rule 1* is applicable on the instance. Since there are $O(mn)$ edges in the instance, applying Reduction Rule 1* on $(F_1,F_2,\ldots,F_m;k)$ until it is unapplicable takes time $O(m^3 n^5)$.

All the other analysis for the Branching Rules about the {\sc sMaf} problem is similar to that about the {\sc hMaf} problem.
Since among all recurrence relations of these branching rules for the {\sc sMaf} problem, the worst one is that of Branching Rule 3*, $T(k) \leq 2T(k-1)+2T(k-2)$, the time complexity of the algorithm Alg-{\sc sMaf} is $O(2.74^k m^3n^5)$.
\end{proof}
\end{theorem}

\section{Conclusion}

In this paper, we studied two versions of the Maximum Agreement Forest problem on multiple rooted multifurcating phylogenetic trees. For the hard version (the {\sc hMaf} problem), we presented the first parameterized algorithm with running time $O(2.42^k m^3 n^4)$; for the soft version (the {\sc sMaf} problem), we presented the first parameterized algorithm with running time $O(2.74^k m^3 n^5)$.

It is relatively simple to develop parameterized algorithms of running time $O^*(3^k)$
for the {\sc hMaf} problem by combining the techniques used in~\cite{14} and~\cite{19},
which also uses a branch-and-bound scheme that has been used in most previous parameterized
algorithms for the Maximum Agreement Forest problem: removing edges in all trees but
branching only on a fixed tree. However, achieving improvements on the algorithm complexity
by simple modifications of the scheme does not seem to be easy.

Thus in the current paper, we proposed a new branch-and-bound scheme where branching
operations can be applied on different trees. The difficulty we had to overcome for
designing the new scheme was how to ensure that each branching operation could effectively
influence the value of the parameter. When the instance under consideration satisfies the
label-set isomorphism property, we had been able to show that branching on different
$X$-forests in the instance is feasible. For the case where the instance under consideration
does not satisfy the label-set isomorphism property, we presented the $m$-BR-process,
and successfully proved that during the $m$-BR-process, the executions of Branching
Rule 1 on different $X$-forests would also effectively influence the value of the parameter.

Although the time
complexity of the best algorithm~\cite{zhizhongchen} for the Maximum Agreement Forest
problem on two rooted binary phylogenetic trees, which is $O(2.344^k n)$, is better
than that of our algorithm Alg-{\sc hMaf}, the methods of the algorithm in~\cite{zhizhongchen} seem
difficult to extend to solving the {\sc hMaf} problem.

To solve the {\sc sMaf} problem, we extended the $m$-BR-process to the $m$-BR*-process, and successfully presented a parameterized algorithm for it with running time $O(2.74^k m^3 n^5)$.
It should be remarked that the soft version of the {\sc Maf} problem is more complicated
than the hard version of the problem, and that constructing an MAF for more than two
$X$-forests for the soft version of the problem is much more complicated than that for only
two $X$-forests. It seems not easy to get an algorithm for the soft version of the {\sc Maf}
problem by simply extending the techniques presented by Whidden in~\cite{17}, who gave an
algorithm of running time $O(2.42^k n)$ for the Maximum Agreement Forest problem on two
rooted multifurcating phylogenetic trees in which all polytomies are soft.

We believe that our new schemes, the $m$-BR-process and its extension $m$-BR*-process, will have further applications in the study of approximation algorithms and parameterized algorithms for the Maximum Agreement
Forest problem on two or more phylogenetic trees. Thus, it would be an interesting direction for future research. Another interesting direction for future research is improving the complexities of our algorithms. However, such an improvement seems to require new
observations in the graph structures of phylogenetic trees and new algorithmic techniques.

\bibliography{mybibfile}

\begin{thebibliography}{10}
\expandafter\ifx\csname url\endcsname\relax
  \def\url#1{\texttt{#1}}\fi
\expandafter\ifx\csname urlprefix\endcsname\relax\def\urlprefix{URL }\fi
\expandafter\ifx\csname href\endcsname\relax
  \def\href#1#2{#2} \def\path#1{#1}\fi

\bibitem{1}
D.~Robinson, L.~Foulds, Comparison of phylogenetic trees, Mathematical
  Biosciences 53~(1--2) (1981) 131--147.

\bibitem{2}
M.~Li, J.~Tromp, L.~Zhang, On the nearest neighbour interchange distance
  between evolutionary trees, Journal on Theoretical Biology 182~(4) (1996)
  463--467.

\bibitem{hybridization}
M.~Baroni, S.~Grunewald, V.~Moulton, C.~Semple, Bounding the number of
  hybridisation events for a consistent evolutionary history, Journal of
  Mathematical Biology 51~(2) (2005) 171--182.

\bibitem{3}
P.~Buneman, The recovery of trees from measures of dissimilarity, in:
  D.~Kendall, P.~Tautu (Eds.), Mathematics in the Archeological and Historical
  Sciences, Edinburgh University Press, 1971, pp. 387--395.

\bibitem{4}
D.~L. Swofford, G.~J. Olsen, P.~J. Waddell, D.~M. Hillis, {Phylogenetic
  inference}, in: E.~Hillis, C.~Moritz, B.~Mable (Eds.), Molecular Systematics,
  2nd ed., 1996, pp. 407--514.

\bibitem{dudas}
G.~Dudas, T.~Bedford, S.~Lycett, A.~Rambaut, Reassortment between influenza {B}
  lineages and the emergence of a coadapted {PB1-PB2-HA} gene complex,
  Molecular Biology and Evolution 32~(1) (2014) 162--172.

\bibitem{beiko}
R.~G. Beiko, N.~Hamilton, Phylogenetic identification of lateral genetic
  transfer events, BMC Evolutionary Biology 6~(1) (2006) 15.

\bibitem{whidden5}
C.~Whidden, N.~Zeh, R.~Beiko, Supertrees based on the subtree prune-and-regraft
  distance, Systematic Biology 63~(4) (2014) 566--581.

\bibitem{whidden6}
C.~Whidden, A.~Frederick, I.~Matsen, Quantifying {MCMC} exploration of
  phylogenetic tree space, Systematic Biology 64~(3) (2015) 472.

\bibitem{6}
B.~Allen, M.~Steel, Subtree transfer operations and their induced metrics on
  evolutionary trees, Annals of Combinatorics 5~(1) (2001) 1--15.

\bibitem{10}
M.~Bordewich, C.~McCartin, C.~Semple, A 3-approximation algorithm for the
  subtree distance between phylogenies, Journal of Discrete Algorithms 6~(3)
  (2008) 458--471.

\bibitem{5}
J.~Hein, T.~Jiang, L.~Wang, K.~Zhang, On the complexity of comparing
  evolutionary trees, Discrete Applied Mathematics 71 (1996) 153--169.

\bibitem{7}
M.~Bordewich, C.~Semple, On the computational complexity of the rooted subtree
  prune and regraft distance, Annals of Combinatorics 8~(4) (2005) 409--423.

\bibitem{21}
J.~Felsenstein, Phylogenies and the comparative method, American Naturalist
  (1985) 1--15.

\bibitem{22}
A.~Grafen, The phylogenetic regression, Philosophical Transactions of the Royal
  Society of London. Series B, Biological Sciences 326~(1233) (1989) 119--157.

\bibitem{27}
J.~Fehrer, B.~Gemeinholzer, J.~Chrtek, S.~Br{\"a}utigam, Incongruent plastid
  and nuclear {DNA} phylogenies reveal ancient intergeneric hybridization in
  {P}ilosella hawkweeds ({H}ieracium, {C}ichorieae, {A}steraceae), Molecular
  Phylogenetics and Evolution 42~(2) (2007) 347--361.

\bibitem{28}
O.~Paun, C.~Lehnebach, J.~T. Johansson, P.~Lockhart, E.~H{\"o}randl,
  Phylogenetic relationships and biogeography of {R}anunculus and allied genera
  ({R}anunculaceae) in the {M}editerranean region and in the {E}uropean
  {A}lpine {S}ystem, Taxon 54~(4) (2005) 911--932.

\bibitem{23}
W.~Maddison, Reconstructing character evolution on polytomous cladograms,
  Cladistics 5~(4) (1989) 365--377.

\bibitem{24}
J.~A. Coyne, S.~Elwyn, S.~Y. Kim, A.~Llopart, Genetic studies of two sister
  species in the drosophila melanogaster subgroup, {D}. yakuba and {D}.
  santomea, Genetical Research 84~(01) (2004) 11--26.

\bibitem{25}
R.~M. Kliman, P.~Andolfatto, J.~A. Coyne, F.~Depaulis, M.~Kreitman, A.~J.
  Berry, J.~McCarter, J.~Wakeley, J.~Hey, The population genetics of the origin
  and divergence of the drosophila simulans complex species, Genetics 156~(4)
  (2000) 1913--1931.

\bibitem{26}
K.~Takahashi, Y.~Terai, M.~Nishida, N.~Okada, Phylogenetic relationships and
  ancient incomplete lineage sorting among cichlid fishes in lake tanganyika as
  revealed by analysis of the insertion of retroposons, Molecular Biology and
  Evolution 18~(11) (2001) 2057--2066.

\bibitem{17}
C.~Whidden, R.~G. Beiko, N.~Zeh, Fixed-parameter and approximation algorithms
  for maximum agreement forests of multifurcating trees, Algorithmica 74~(3)
  (2016) 1019--1054.

\bibitem{14}
J.~Chen, J.-H. Fan, S.-H. Sze, Parameterized and approximation algorithms for
  maximum agreement forest in multifurcating trees, Theoretical Computer
  Science 562 (2015) 496--512.

\bibitem{18}
Z.-Z. Chen, L.~Wang, Algorithms for reticulate networks of multiple
  phylogenetic trees, IEEE/ACM Transactions on Computational Biology and
  Bioinformatics (TCBB) 9~(2) (2012) 372--384.

\bibitem{leo2013}
L.~Van~Iersel, S.~Linz, A quadratic kernel for computing the hybridization
  number of multiple trees, Information Processing Letters 113~(9) (2013)
  318--323.

\bibitem{8}
E.~M. Rodrigues, M.-F. Sagot, Y.~Wakabayashi, Some approximation results for
  the maximum agreement forest problem, in: Approximation, Randomization, and
  Combinatorial Optimization: Algorithms and Techniques, Springer, 2001, pp.
  159--169.

\bibitem{9}
M.~L. Bonet, K.~S. John, R.~Mahindru, N.~Amenta, Approximating subtree
  distances between phylogenies, Journal of Computational Biology 13~(8) (2006)
  1419--1434.

\bibitem{11}
E.~M. Rodrigues, M.-F. Sagot, Y.~Wakabayashi, The maximum agreement forest
  problem: approximation algorithms and computational experiments, Theoretical
  Computer Science 374~(1) (2007) 91--110.

\bibitem{12}
C.~Whidden, N.~Zeh, A unifying view on approximation and {FPT} of agreement
  forests, Algorithms in Bioinformatics 5724 (2009) 390--401.

\bibitem{shiapprox}
F.~Shi, Q.~Feng, J.~You, J.~Wang, Improved approximation algorithm for maximum
  agreement forest of two rooted binary phylogenetic trees, Journal of
  Combinatorial Optimization 32~(1) (2016) 111--143.

\bibitem{Schalekamp}
F.~Schalekamp, A.~V. Zuylen, S.~V.~D. Ster, A duality based 2-approximation
  algorithm for maximum agreement forest, in: 43rd International Colloquium on
  Automata, Languages, and Programming (ICALP 2016), Schloss
  Dagstuhl-Leibniz-Zentrum fuer Informatik, 2016, pp. 70:1--70:14.

\bibitem{33}
L.~Van~Iersel, S.~Kelk, N.~Lekic, L.~Stougie, Approximation algorithms for
  nonbinary agreement forests, SIAM Journal on Discrete Mathematics 28~(1)
  (2014) 49--66.

\bibitem{15}
F.~Chataigner, Approximating the maximum agreement forest on k trees,
  Information Processing Letters 93~(5) (2005) 239--244.

\bibitem{Asish}
A.~Mukhopadhyay, P.~Bhabak, A 3-factor approximation algorithm for a minimum
  acyclic agreement forest on k rooted, binary phylogenetic trees, arXiv
  preprint arXiv:1407.7125.

\bibitem{20}
J.~Chen, F.~Shi, J.~Wang, Approximating maximum agreement forest on multiple
  binary trees, Algorithmica (2015) 1--23.

\bibitem{fptbook}
R.~G. Downey, M.~R. Fellows, Parameterized Complexity, New York, U.S.:
  Springer, 1999.

\bibitem{siam}
C.~Whidden, R.~G. Beiko, N.~Zeh, Fixed-parameter algorithms for maximum
  agreement forests, SIAM Journal on Computing 42~(4) (2013) 1431--1466.

\bibitem{zhizhongchen}
Z.-Z. Chen, Y.~Fan, L.~Wang, Faster exact computation of r{SPR} distance,
  Journal of Combinatorial Optimization 29~(3) (2013) 605--635.

\bibitem{Yang}
F.~Shi, J.~Wang, Y.~Yang, Q.~Feng, W.~Li, J.~Chen, A fixed-parameter algorithm
  for the maximum agreement forest problem on multifurcating trees, Science
  China Information Sciences 59~(1) (2016) 1--14.

\bibitem{19}
F.~Shi, J.~Wang, J.~Chen, Q.~Feng, J.~Guo, Algorithms for parameterized maximum
  agreement forest problem on multiple trees, Theoretical Computer Science 554
  (2014) 207--216.

\bibitem{Hallett}
M.~Hallett, C.~McCartin, A faster {FPT} algorithm for the maximum agreement
  forest problem, Theory of Computing Systems 41~(3) (2007) 539--550.

\bibitem{vcold}
J.~Chen, I.~A. Kanj, W.~Jia, Vertex cover: further observations and further
  improvements, Journal of Algorithms 41 (2001) 280--301.

\end{thebibliography}

\end{document}